\documentclass[floats,floatfix,showpacs,amssymb,prd,superscriptaddress,twocolumn,aps]{revtex4}
\usepackage{graphicx, epsfig, amssymb} %include figure files
\usepackage{amsmath, amsfonts}
\usepackage{gensymb}
\usepackage{bm} %include bold math: \bm{} creates bold letters in math mode

\usepackage[linktocpage]{hyperref}
\usepackage[caption=false]{subfig}
\usepackage[usenames]{color}

\def\be{\begin{equation}}
\def\ee{\end{equation}}
\def\beq{\begin{eqnarray}}
\def\eeq{\end{eqnarray}}

\newcommand{\bea}{\begin{eqnarray}}
\newcommand{\eea}{\end{eqnarray}}
\newcommand{\ben}{\begin{enumerate}}
\newcommand{\een}{\end{enumerate}}
\newcommand{\bi}{\begin{itemize}}
\newcommand{\ei}{\end{itemize}}

\pdfoutput=1

\begin{document}

%%%%%%%%%%%%%%%%%
%%%   TITLE   %%%
%%%%%%%%%%%%%%%%%
\title{Shadows of Kerr black holes with scalar hair}

 \author{Pedro V. P. Cunha}
  \affiliation{Departamento de F\'\i sica, Universidade de Coimbra, 3004-516 Coimbra, Portugal}
   \affiliation{
   Departamento de F\'\i sica da Universidade de Aveiro and CIDMA, 
   Campus de Santiago, 3810-183 Aveiro, Portugal.
 }

 \author{Carlos~A.~R.~Herdeiro}
   \affiliation{
   Departamento de F\'\i sica da Universidade de Aveiro and CIDMA, 
   Campus de Santiago, 3810-183 Aveiro, Portugal.
 }

 \author{Eugen Radu}
   \affiliation{
   Departamento de F\'\i sica da Universidade de Aveiro and CIDMA, 
   Campus de Santiago, 3810-183 Aveiro, Portugal.
 }
 
  \author{Helgi F. R\'unarsson}
   \affiliation{
   Departamento de F\'\i sica da Universidade de Aveiro and CIDMA, 
   Campus de Santiago, 3810-183 Aveiro, Portugal.
 }

%%%%%%%%%%%%%%%%
%%%   DATE   %%%
%%%%%%%%%%%%%%%%

\date{October 2015}

%%%%%%%%%%%%%%%%%%%%
%%%   ABSTRACT   %%%
%%%%%%%%%%%%%%%%%%%%
\begin{abstract}
Using backwards ray tracing, we study the shadows of Kerr black holes with scalar hair (KBHsSH). KBHsSH interpolate continuously between Kerr BHs and boson stars (BSs), so we start by investigating the lensing of light due to BSs. Moving from the weak to the strong gravity region, BSs  - which by themselves have no shadows - are classified, according to the lensing produced, as: $(i)$ \textit{non-compact}, which yield no multiple images; $(ii)$ \textit{compact}, which produce an increasing number of Einstein rings and multiple images of the whole celestial sphere; $(iii)$ \textit{ultra-compact}, which possess light rings, yielding an infinite number of images with (we conjecture) a self-similar structure. The shadows of KBHsSH, for \textit{Kerr-like horizons} and non-compact BS-like hair, are analogous to, but distinguishable from, those of comparable Kerr BHs. But for \textit{non-Kerr-like horizons} and ultra-compact BS-like hair, the shadows of KBHsSH are drastically different: novel shapes arise, sizes are considerably smaller and multiple shadows of a \textit{single} BH become possible. Thus, KBHsSH provide quantitatively and qualitatively new templates for ongoing (and future) very large baseline interferometry (VLBI) observations of BH shadows, such as those of the Event Horizon Telescope.
\end{abstract}

%%%%%%%%%%%%%%%%
%%%   PACS   %%%
%%%%%%%%%%%%%%%%

\pacs{
04.20.-q, % classical general relativity
04.70.Bw  % classical black holes
04.80.Cc 	% Experimental tests of gravitational theories
}

%%%%%%%%%%%%%%%%%%%%%%
%%%   MAKE TITLE   %%%
%%%%%%%%%%%%%%%%%%%%%%

\maketitle
%%%%%%%%%%%%%%%%%%%%%%%%%%%%%%%%%%%%%%%%%%%%%%%%%%%%%%%%%%%%%%%%%%%%%
%%%%%%%%%%%%%%%%%%%%%%%%%%%%%%%%%%%%%%%%%%%%%%%%%%%%%%%%%%%%%%%%%%%%%
\noindent{\bf {\em Introduction.}} 100 years after General Relativity (GR) was formulated, we finally face a realistic prospect of testing one of its most dramatic consequences: black holes (BHs). The evidence for astrophysical BHs, gathered for over half a century, has built a strong case~\cite{Narayan:2013gca}, but it could not confirm the existence of event horizons, the defining property of BHs.  The near future promises to open up a new channel of observation - gravitational waves - and deliver electromagnetic measurements of unprecedented precision, hopefully clarifying this central issue~\cite{Berti:2015itd}.

A particularly exciting prospect is the use of VLBI techniques to resolve the angular scale of the event horizon for some supermassive BH candidates and determine the corresponding \textit{BH shadow}~\cite{1973blho.conf..215B,Falcke:1999pj}. Its observation would probe the spacetime geometry in the vicinity of the horizon and consequently test the existence and properties of the latter~\cite{Loeb:2013lfa}. It is therefore timely to study BH models that yield phenomenological deviations from the paradigmatic GR BH, described by the Kerr metric.

One approach is to parameterize families of metric deviations from Kerr~\cite{Johannsen:2010ru,Johannsen:2011dh,Johannsen:2015pca,Johannsen:2013rqa,Cardoso:2014rha,Rezzolla:2014mua}. Another approach is to use exact solutions of GR (or generalizations thereof)  yielding  deviations from Kerr ($e.g.$~\cite{Amarilla:2011fx,Grenzebach:2015oea,Moffat:2015kva}). Exact solutions with physically reasonable and astrophysically plausible matter sources, however, are scarce; but Kerr BHs with scalar hair (KBHsSH)~\cite{Herdeiro:2014goa} are arguably one such model. These are exact solutions of Einstein's gravity minimally coupled to a massive complex scalar field, and interpolate between Kerr BHs and gravitating solitons - \textit{boson stars} (BSs)~\cite{Schunck:2003kk} - suggested as dark matter candidates (in the Newtonian limit) and BH mimickers~\cite{AmaroSeoane:2010qx,Li:2013nal,Suarez:2013iw}. 

In this letter we show that the shadows of KBHsSH are distinguishable, or even drastically different, from those of Kerr BHs, and can thus yield new templates for the ongoing and future VLBI searches~\cite{2009astro2010S..68D} of BH shadows.

\begin{figure}[h!]
\begin{center}
\includegraphics[width=0.454\textwidth]{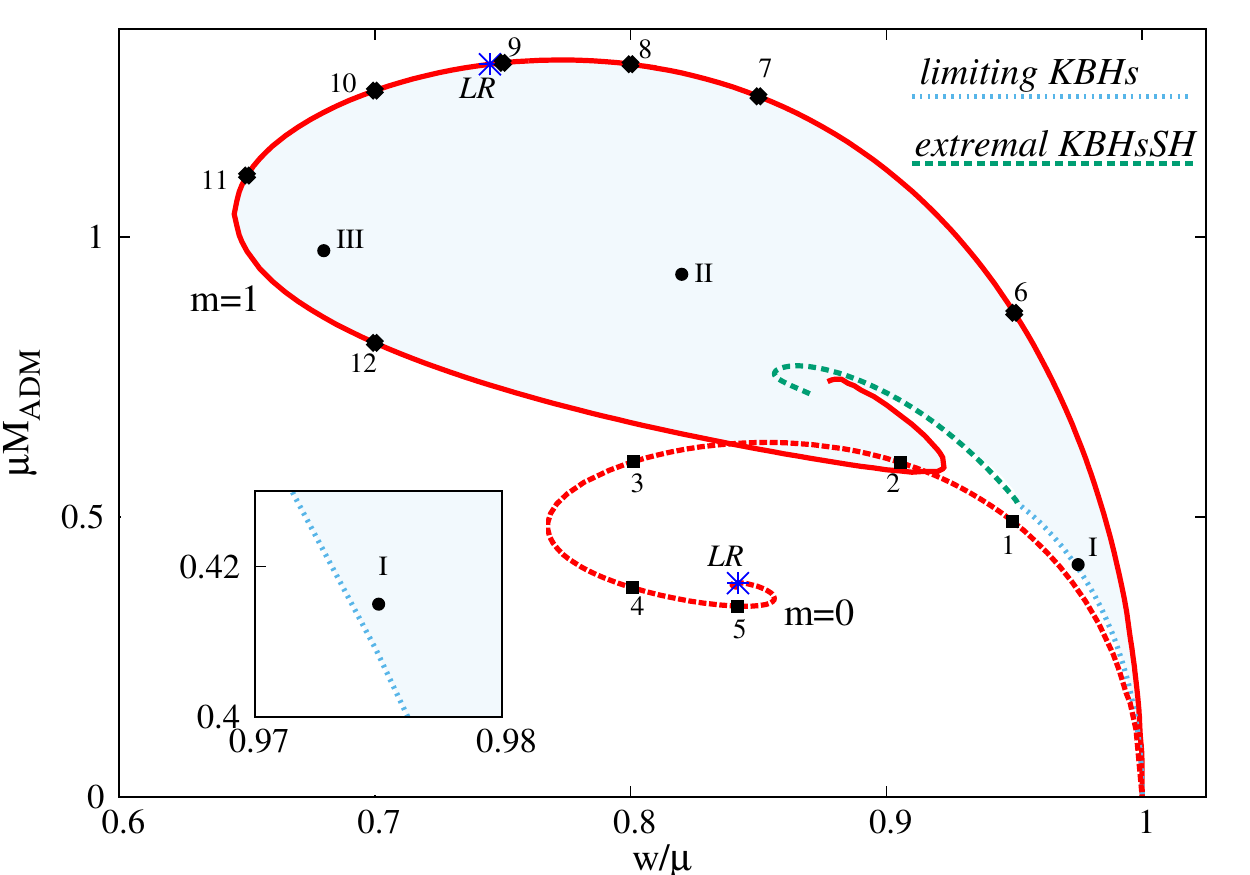}
\caption{BS solutions with $m=0,1$ (red dashed and solid lines) in an ADM mass $M_{\rm ADM}$ $vs.$ scalar field frequency $w$ diagram (in units of $\mu$). For $m=1$, KBHsSH exist within the shaded region. Points 1-12 (I-III) correspond to the BSs (KBHsSH) exhibited below. LR=light ring.}
\label{spaceofsolutions}
\end{center}
\end{figure}

\noindent{\bf {\em The solutions.}} KBHsSH can be expressed by a stationary and axi-symmetric line element, in spheroidal coordinates $(t,r,\theta,\phi)$~\cite{Herdeiro:2014goa,Herdeiro:2015gia}, together with the (mass $\mu$) scalar field $\Psi=\phi(r,\theta) \, e^{i(m\varphi-w t)}$, where $w$ is the frequency and $m\in \mathbb{Z}^+$ is the azimuthal harmonic index.  The metric functions and $\phi$ are determined numerically by solving five coupled, non-linear PDEs~\cite{Herdeiro:2015gia}. For $m\geqslant 1$, both BSs and KBHsSH solutions can be obtained. For $m=0$, spherical BSs exist. The space of solutions for $m=0,1$ is summarized in Fig.~\ref{spaceofsolutions}. BSs exist for a limited range of $w$ along spiraling curves. KBHsSH exist inside an open set, bounded by the $m=1$ BS curve, a set of Kerr BHs and the set of extremal KBHsSH.

\begin{figure}[h!]
\begin{center}
\includegraphics[width=0.236\textwidth]{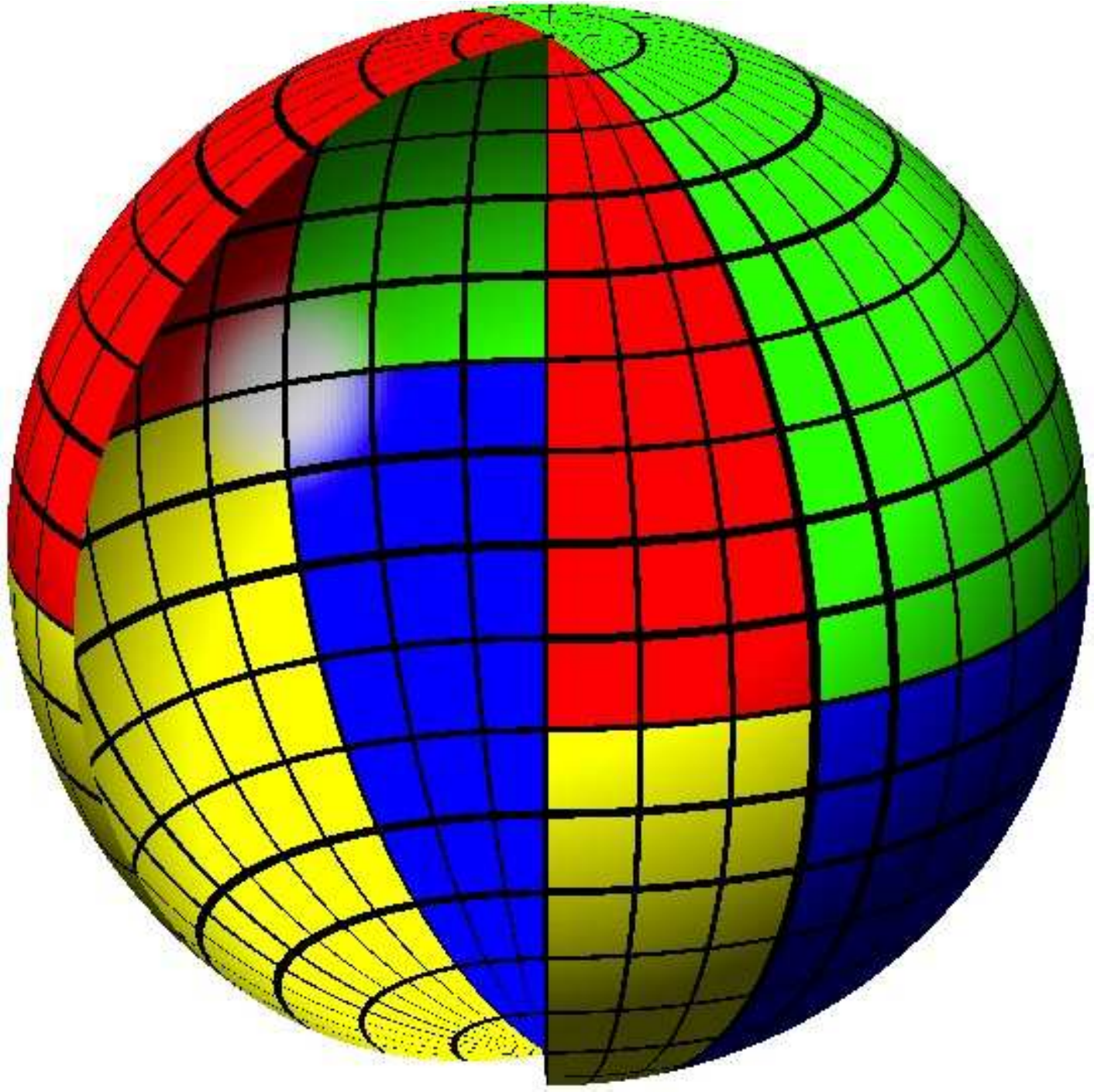}
\includegraphics[width=0.236\textwidth]{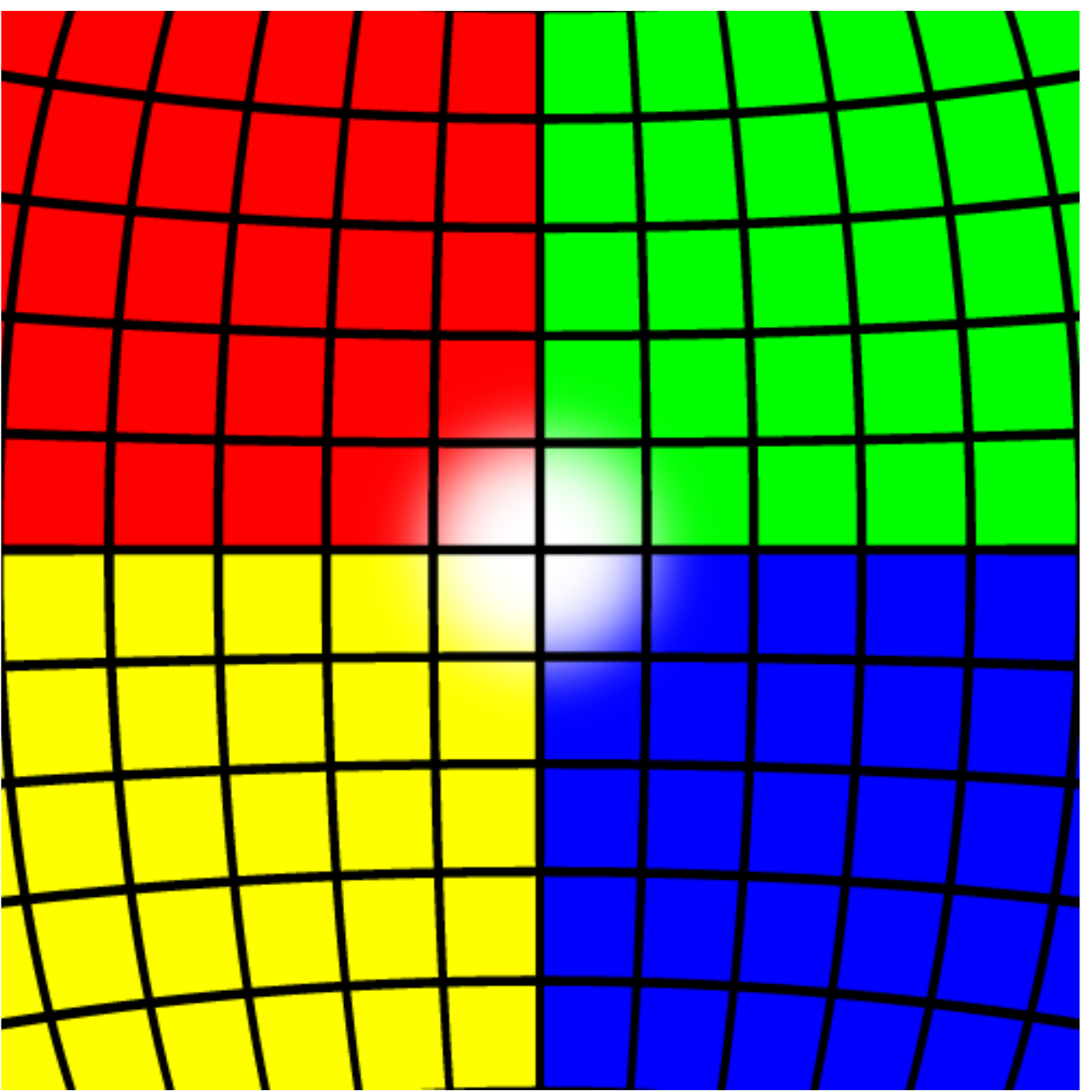}
\caption{(Left) The full celestial sphere. (Right) The viewing angle of the off-centred observer ($\mathcal{O}$).}
\label{background}
\end{center}
\end{figure}

\noindent{\bf {\em Setup.}} Our goal is to compute photon geodesics in the above geometries and obtain, at a given observation point, the distorted apparent sky, due to the gravitational lensing of BSs, as well as the shadow, when a horizon is present. To interpret the patterns obtained, we divide the ``celestial sphere" light source into four quadrants, each painted with a different color. On top of these, a grid of constant longitude and latitude (black) lines is introduced, with adjacent lines separated by $10^\circ$ - Fig.~\ref{background} (left). This setup mimicks closely the one in~\cite{Bohn:2014xxa}.

The observer, henceforth denoted as $\mathcal{O}$, is placed inside the celestial sphere at an off-centered position to be specified below, and its viewing angle captures $\sim100^\circ$ of that sphere - Fig.~\ref{background} (right). The point on the celestial sphere immediately in front of $\mathcal{O}$ (dubbed $F$) lies at the intersection of the four colored quadrants and it is marked with a white dot.

From the $\mathcal{O}$'s position we span its viewing angle by performing, numerically, backwards ray tracing of $2000\times 2000$ photon trajectories. We integrate these null geodesics until they either reach a point on the celestial sphere or they hit the BH horizon (when it is present). The latter directions form the shadow~\footnote{The celestial sphere is centered around the BH/BS and is typically placed at twice the $R$ coordinate of $\mathcal{O}$. $\mathcal{O}$ is on the equatorial plane. See Sec. III in the Supplemental Material for other details.}. $\mathcal{O}$'s image is obtained upon a projection onto its local frame, by a method similar to that in~\cite{Bohn:2014xxa}.

\noindent{\bf {\em Quantitative shadow parameters.}} 
To analyse the shadows obtained below we introduce six parameters, $\{D_{C},D_{x},D_{y},\bar{r},\sigma_r,\sigma_{\textrm{Kerr}}\}$, mostly following~\cite{Johannsen:2015qca}.
Let $\mathcal{O}$'s image be parametrized by the Cartesian coordinates $(x,y)$, obtained from multiplying the observation angles by the circumferential radius $R$~\footnote{The circumferencial radius is defined such that $2\pi R=\oint d\varphi \sqrt{g_{\varphi\varphi}}$, where the metric component $g_{\varphi\varphi}$ is taken at a spacelike slice and on the equatorial plane, and $\partial_\varphi$ is the azimuthal Killing vector field.} at $\mathcal{O}$'s position. The origin of this coordinate system, $O$, points at the (unlensed) white dot on the celestial sphere. The centre of the shadow, $C$, has $x_C=(x_{\textrm{max}}+x_{\textrm{min}})/2,$
where $x_{\textrm{max}}$ and $x_{\textrm{min}}$ ($y_{\textrm{max}}$ and $y_{\textrm{min}}$) are respectively the maximum and minimum abscissae (ordinates) of the shadow's edge. Due to reflection symmetry for observations on the equatorial plane, $y_C=0$. $C$ and $O$ need not coincide; the \textit{displacement}, $D_C\equiv |x_C|$, measures their difference. The \textit{width} and \textit{height} of the shadow are, respectively, $D_{x}\equiv x_{\textrm{max}}-x_{\textrm{min}}$ and $D_{y}\equiv y_{\textrm{max}}-y_{\textrm{min}}$.

A generic point $P$ on the shadow's edge is at a distance $r\equiv ({y_P}^2 + {(x_P-x_C)}^2)^{1/2}$ from $C$. 
Let $\alpha$ be the angle between the line $\overline{CP}$ and the $x$ axis; the \textit{average radius} is $\bar{r}\equiv \int_0^{2\pi}r(\alpha)\,d\alpha/2\pi $ and the \textit{deviation from sphericity} is  $\sigma_r\equiv \left[\int_0^{2\pi}{\Big(r(\alpha)-\bar{r}\Big)}^2d\alpha/2\pi\right]^{1/2}$. Finally,
\begin{equation} \sigma_{\textrm{K}}\equiv \sqrt{\frac{1}{2\pi}\int_0^{2\pi}{\left(\frac{r(\alpha)-r_{\textrm{Kerr}}(\alpha)}{r_{\textrm{Kerr}}(\alpha)}\right)}^2d\alpha}\ ,
\label{deviation}
\end{equation}
is the \textit{relative deviation from a comparable Kerr BH}, either with the same ADM mass and angular momentum, $M_{\rm ADM}$, $J_{\rm ADM}$, or with the same horizon quantities, $M_{\rm H}, J_{\rm H}$ (as long as the Kerr bound is not violated for the comparable Kerr BH). These comparable Kerr BHs are denoted below as Kerr$_{\rm =ADM}$ and Kerr$_{\rm =H}$, respectively. $C$ is made to coincide for both BHs in~(\ref{deviation}).

\begin{figure}[h!]
\begin{center}
\includegraphics[width=0.155\textwidth]{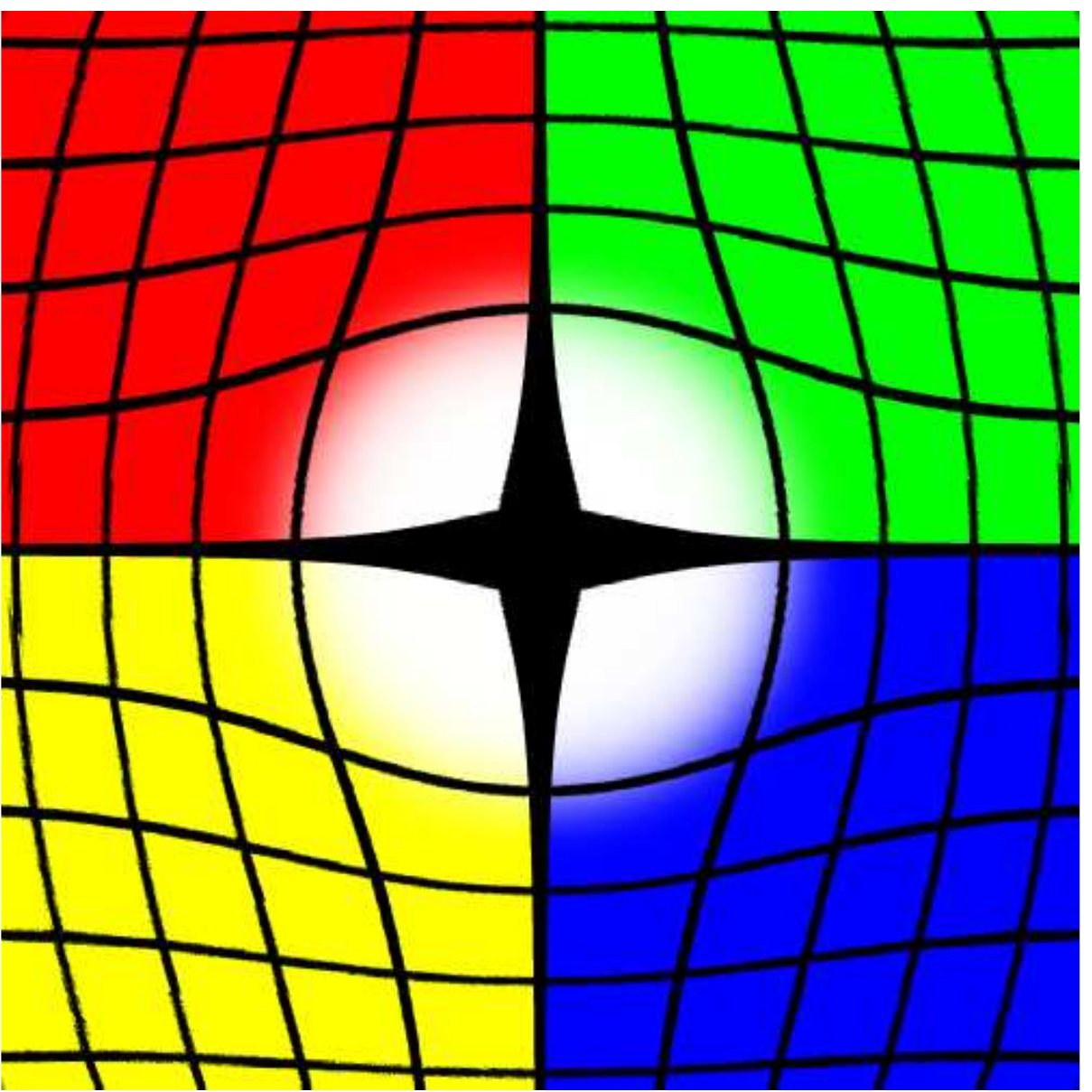}
\includegraphics[width=0.155\textwidth]{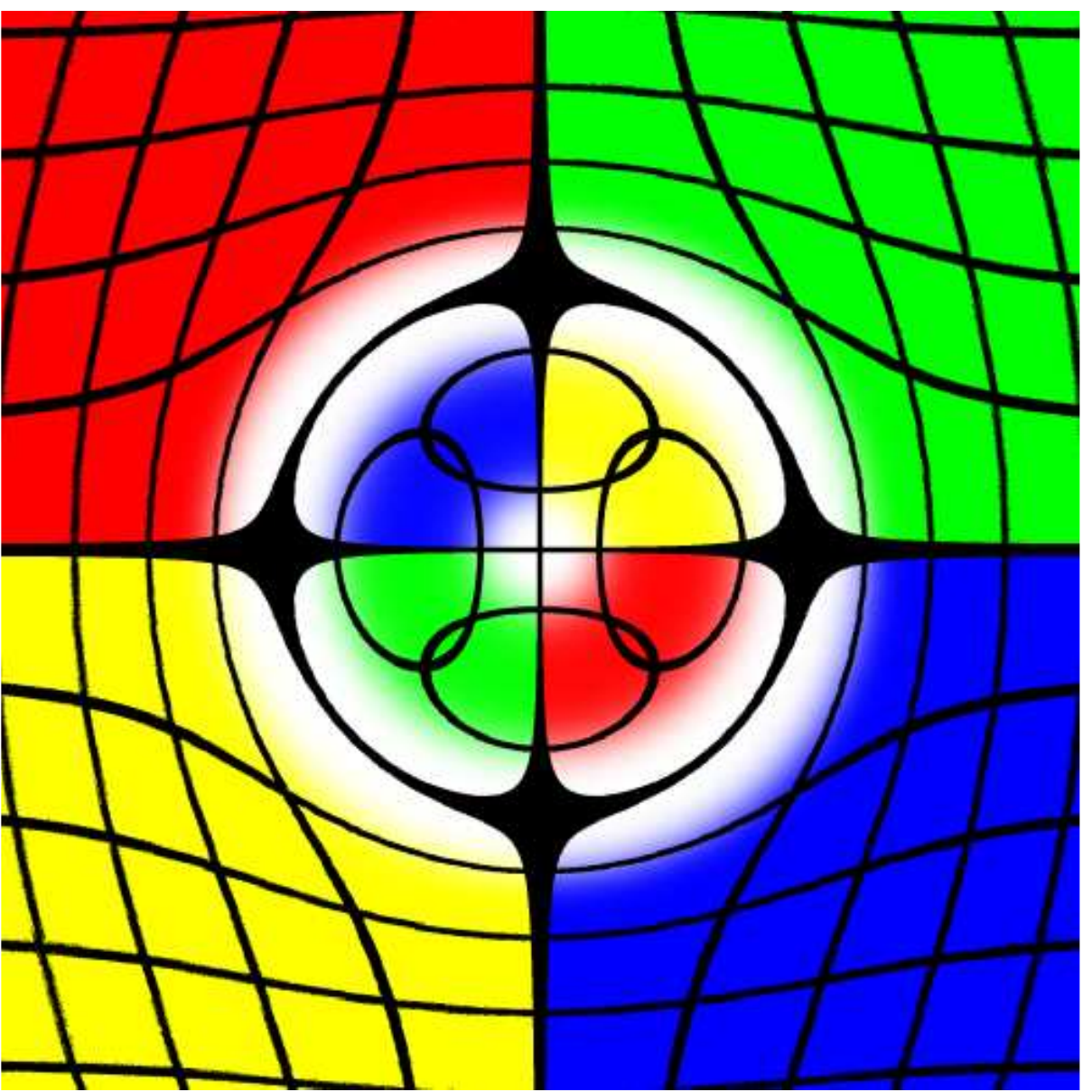}
\includegraphics[width=0.155\textwidth]{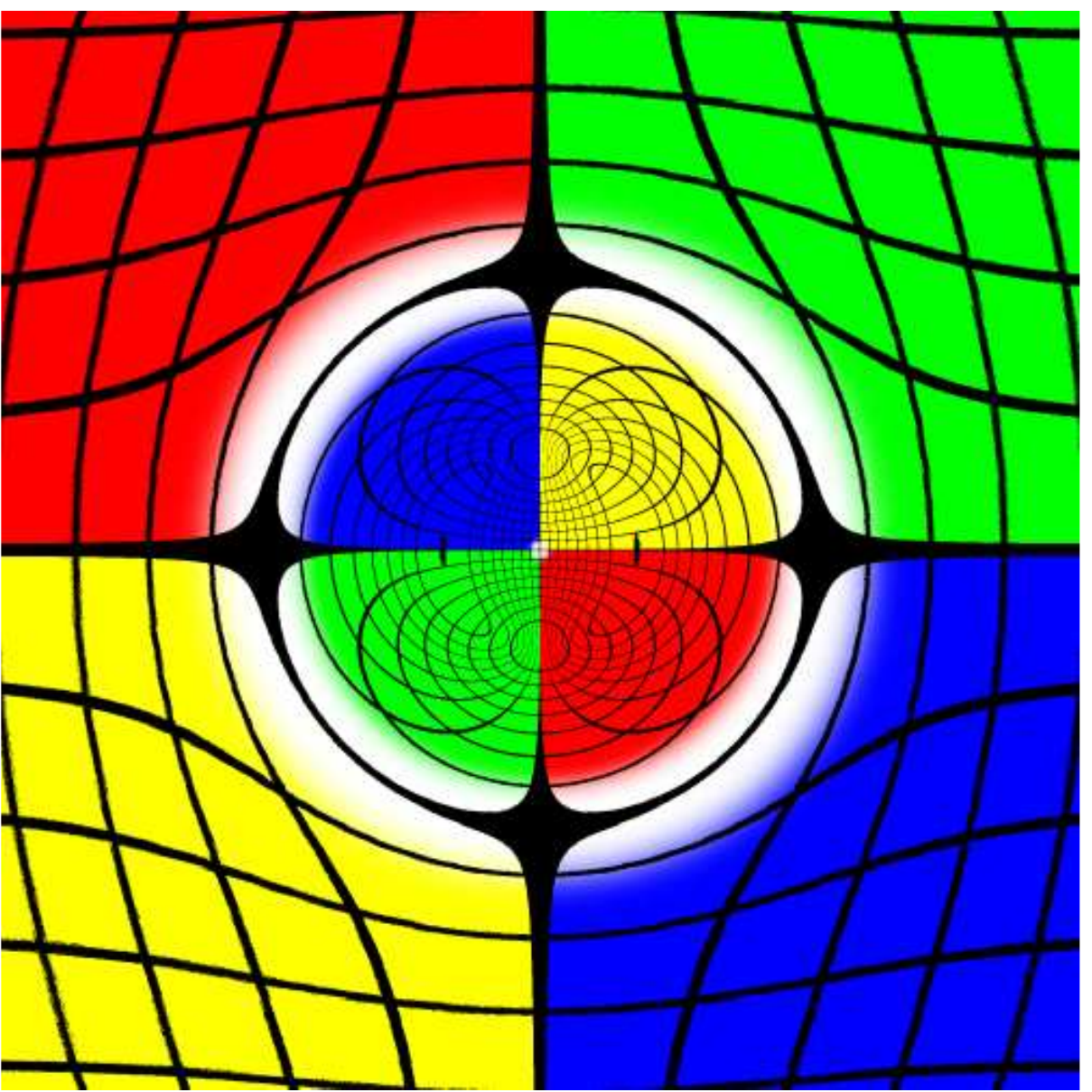}
\includegraphics[width=0.236\textwidth]{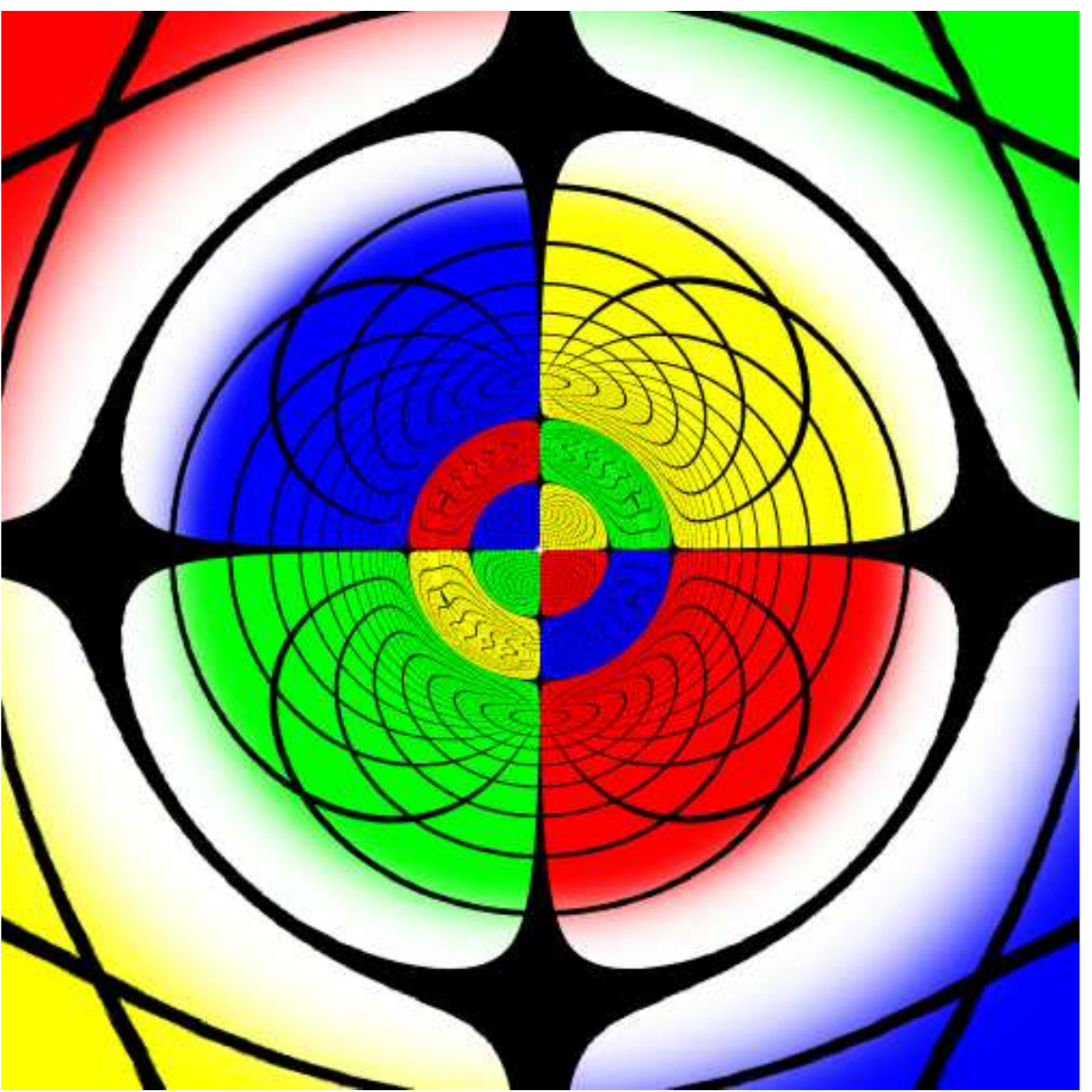}
\includegraphics[width=0.236\textwidth]{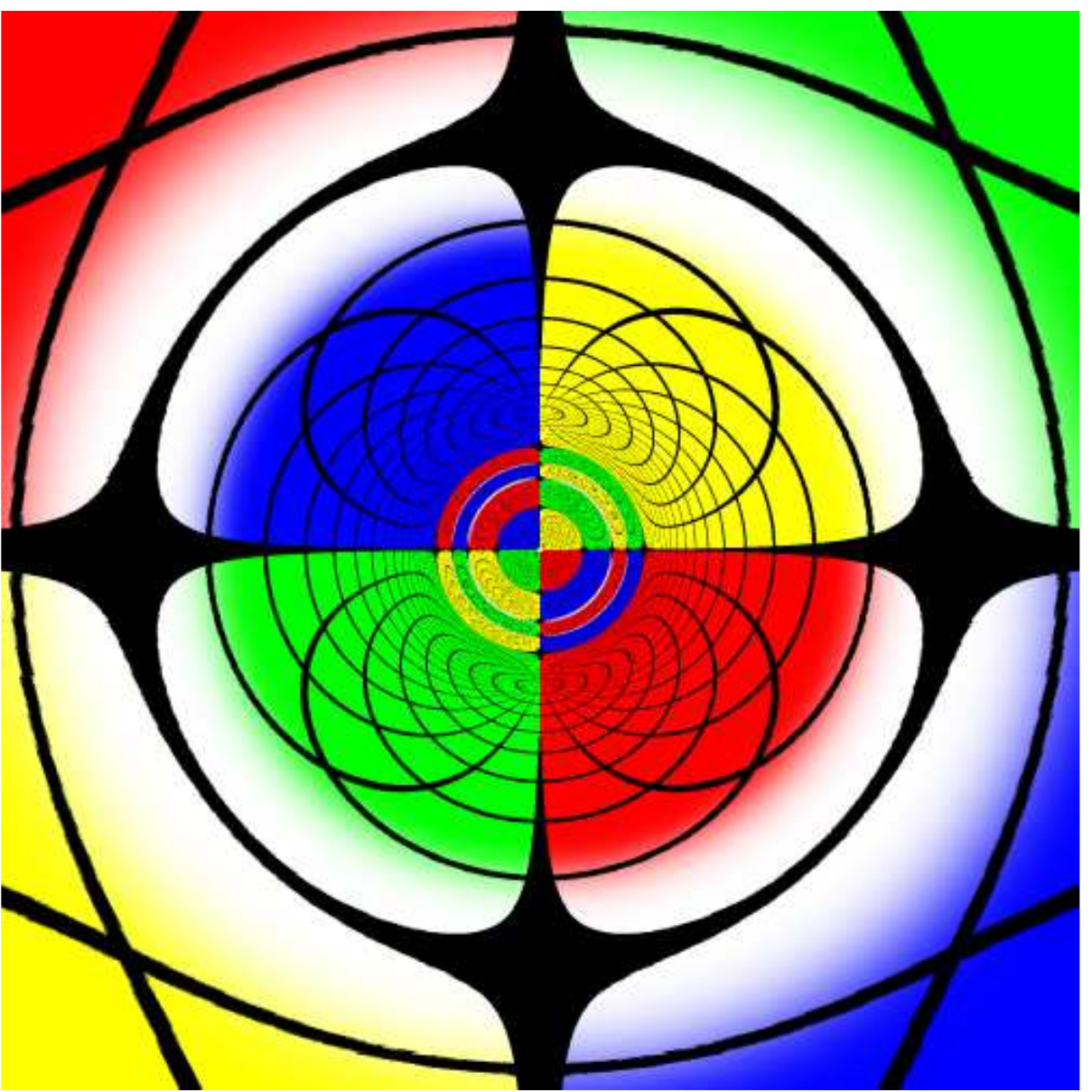}
\caption{Lensing by spherical BSs. From left to right: (top)  $w^{(b1)}_{1,2,3}=0.95; 0.9; 0.8$; (bottom) $w^{(b2)}_{4,5}=0.8; 0.84$ (zoomed).}
\label{SBS}
\end{center}
\end{figure}

\noindent{\bf {\em Lensing by spherical BSs.}}
We first look at the lensing due to spherically symmetric BSs - Fig.~\ref{SBS}. We set $\mathcal{O}$ on the equator and always at $R=22.5/\mu$, for the different BS solutions. Throughout we take $G=1=c$ and in the following, unless otherwise stated, $\mu=1$.

Starting from vacuum ($i.e.$ $w=1$) we find a set of \textit{non-compact} BSs, for which any meaningful effective radius is large as compared to the corresponding Schwarzschild radius. These are ``weak gravity" solutions and the corresponding lensing is illustrated by the BS with $w_1^{(b1)}=0.95$~\footnote{The subscript denotes the number of the solution in Fig.~\ref{spaceofsolutions} or other relevant information: $ER1/ER2$= first/second Einstein ring; $LR$=light ring; $BB1$=first back bending. The superscript denotes the branch. The branch changes after each backbending in the $M$-$w$ plot.} in Fig.~\ref{SBS}, where only a small distortion of the background is observed. Moving further along the spiral, an Einstein ring appears at $w^{(b1)}_{ER1}\simeq0.94$. The Einstein ring is formed by the lensing of $F$, and it encloses two inverted copies of a region around $F$. This is illustrated in Fig.~\ref{SBS} using a BS with $w_2^{(b1)}=0.9$. The appearance of the first Einstein ring defines the transition from non-compact to \textit{compact} BSs. 

Moving further along the spiral the region duplicated inside the Einstein ring becomes an increasingly larger part of the celestial sphere, as can be observed for the BS with $w^{(b1)}_3=0.8$ in Fig.~\ref{SBS}. Shortly  after the spiral's first backbending,  at $w^{(b1\rightarrow b2)}_{BB1}\simeq 0.767$ ($cf.$ Fig.~\ref{spaceofsolutions}), the full celestial sphere becomes duplicated, starting at the BS with $w^{(b2)}_{ER2}\simeq 0.77$. At, and beyond, this point, two further Einstein rings emerge - now corresponding to lensings of the point immediately behind the observer (dubbed $B$). This is illustrated by the bottom left panel in Fig.~\ref{SBS}, for a BS with $w^{(b2)}_4=0.8$. In between these two new Einstein rings, further pairs of Einstein rings can be seen to appear, progressively, further inside the spiral. The first such pair is illustrated in Fig.~\ref{SBS} for $w^{(b2)}_5=0.84$. Each new pair of Einstein rings corresponds to images of either $F$ or $B$, in an alternating fashion, and to a further complete copy of the full celestial sphere.  

An infinite number of copies, and a corresponding self-similar structure, is expected to arise when a light-ring -- corresponding in this case to a photon sphere -- appears~\footnote{In both the spherical and rotating BSs case, it is actually a pair of light rings that appears.}, marking the transition from compact to \textit{ultra-compact}~\cite{Cardoso:2014sna} BSs. This occurs well inside the spiral, on the third branch (after the second backbending), starting at the BS solution with $ w^{(b3)}_{LR}=0.842$, 
marked as the blue star `$LR$' point on the $m=0$ spiral of Fig.~\ref{spaceofsolutions}.

\begin{figure}[h!]
\begin{center}
\includegraphics[width=0.155\textwidth]{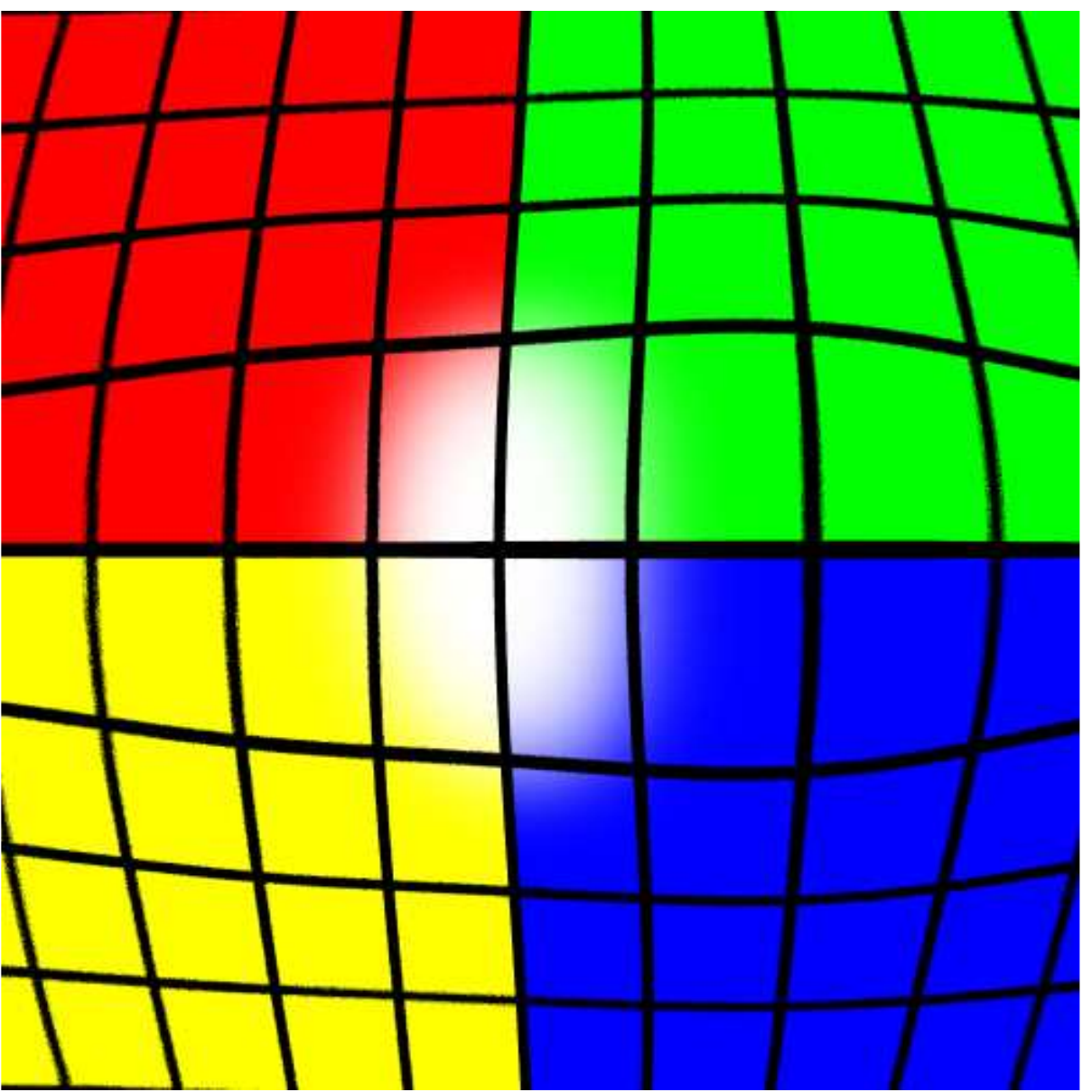}
\includegraphics[width=0.155\textwidth]{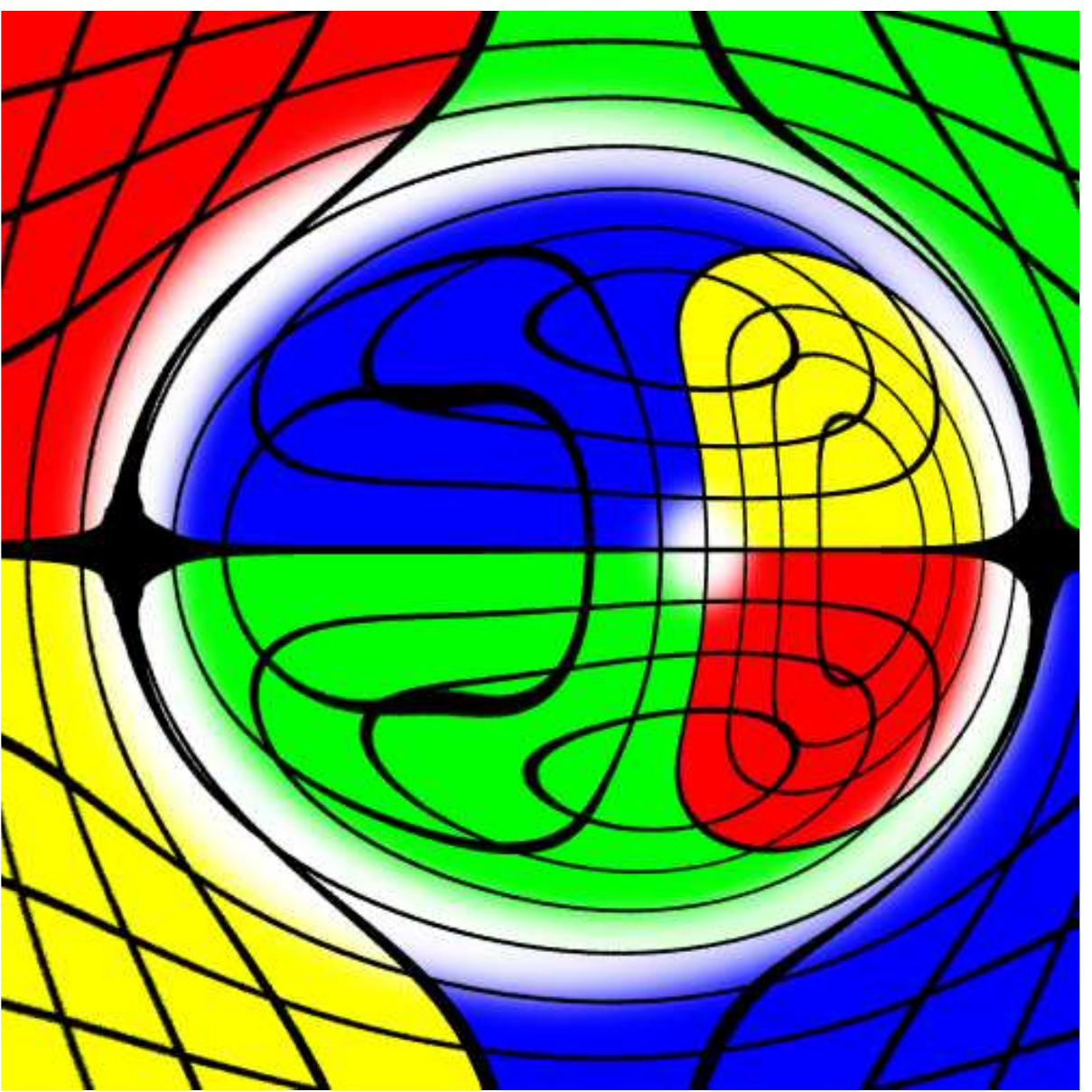}
\includegraphics[width=0.155\textwidth]{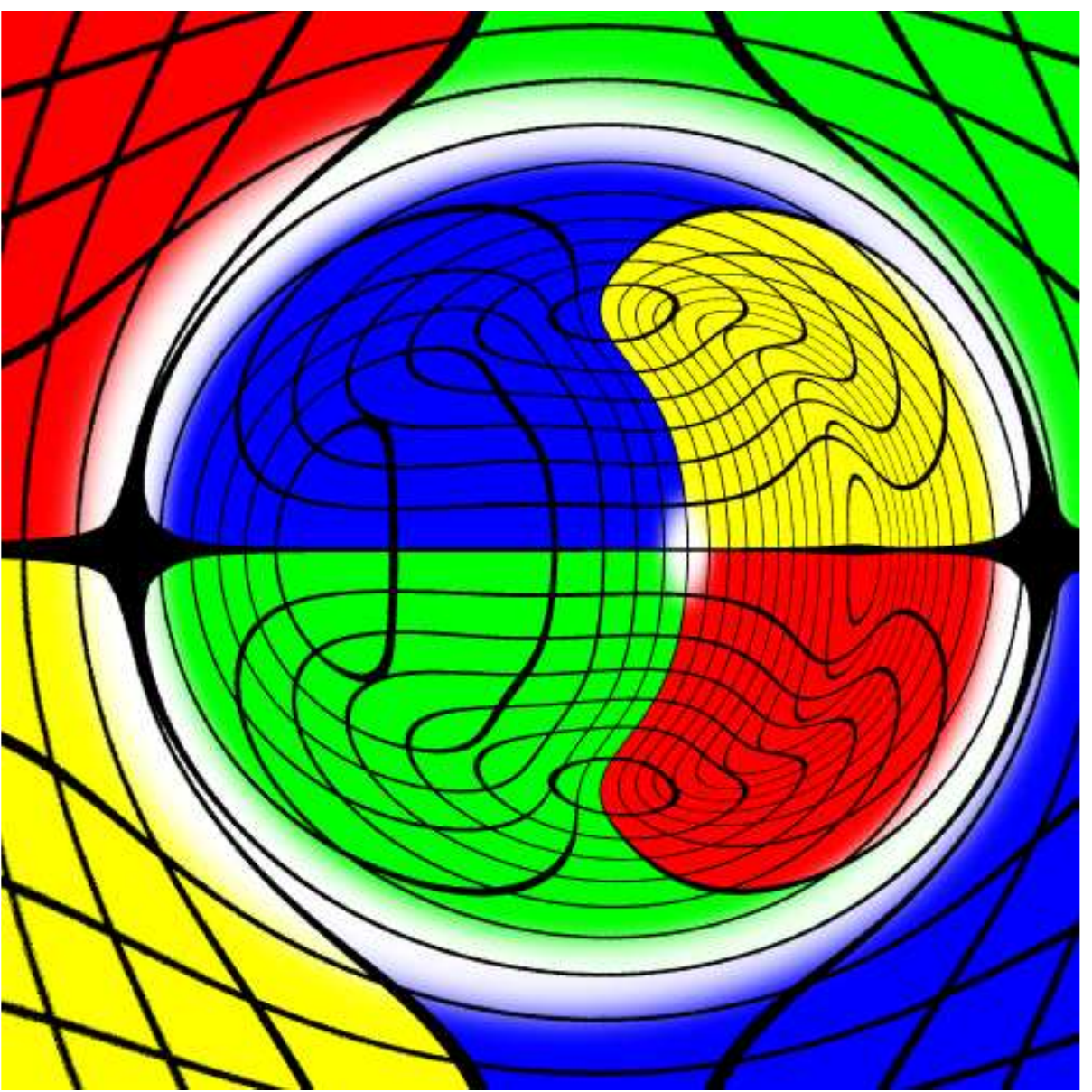}
\includegraphics[width=0.236\textwidth]{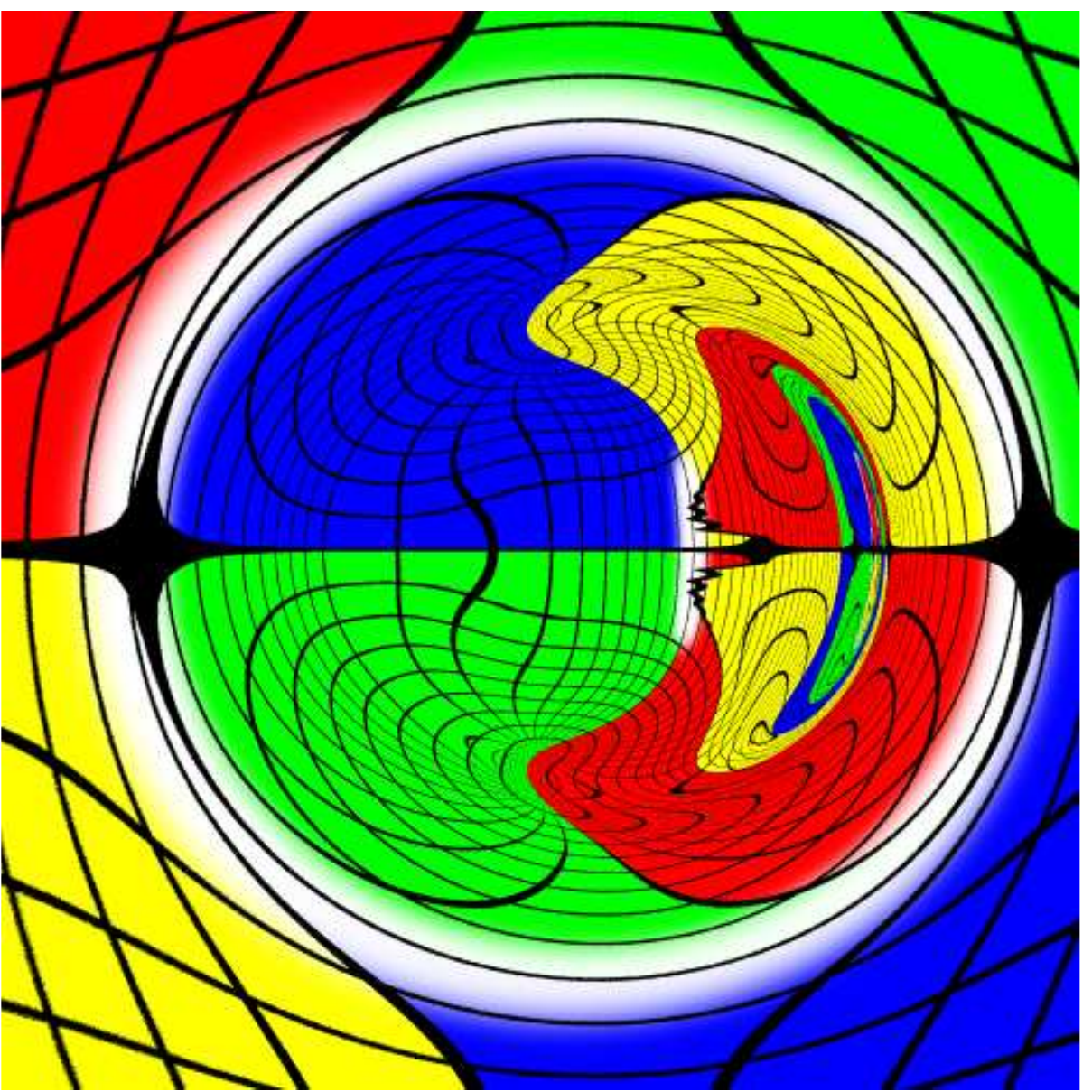}
\includegraphics[width=0.236\textwidth]{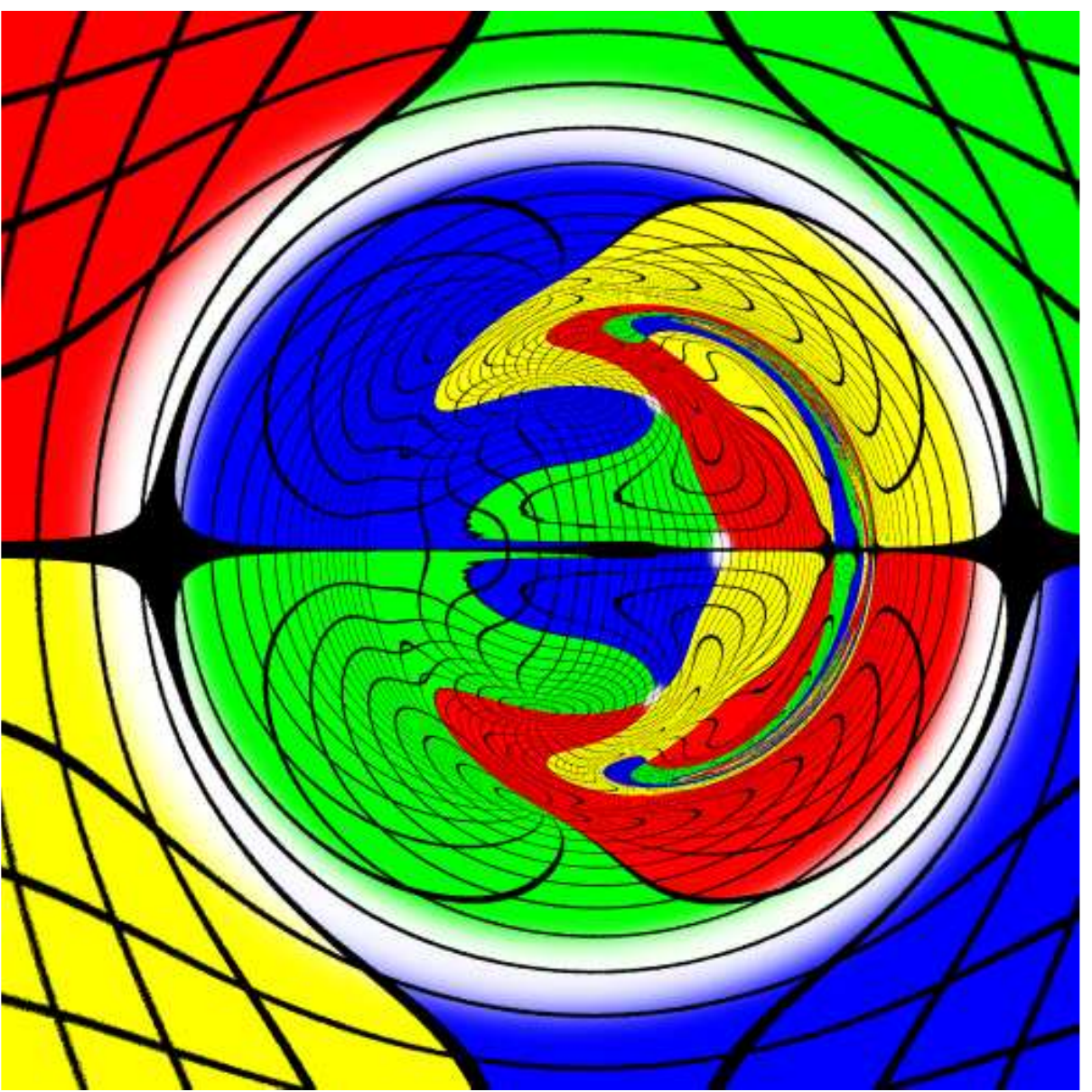}
\includegraphics[width=0.236\textwidth]{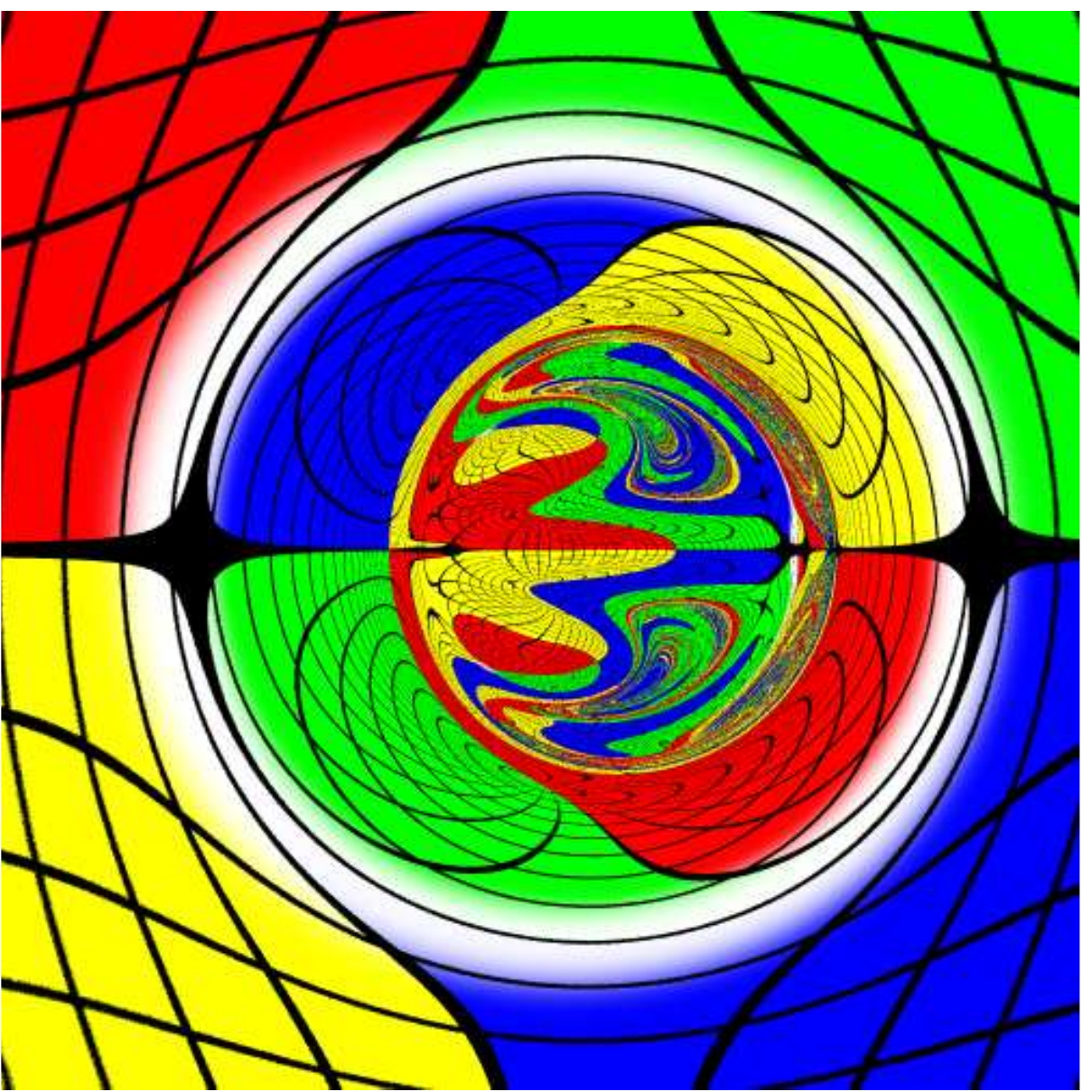}
\includegraphics[width=0.236\textwidth]{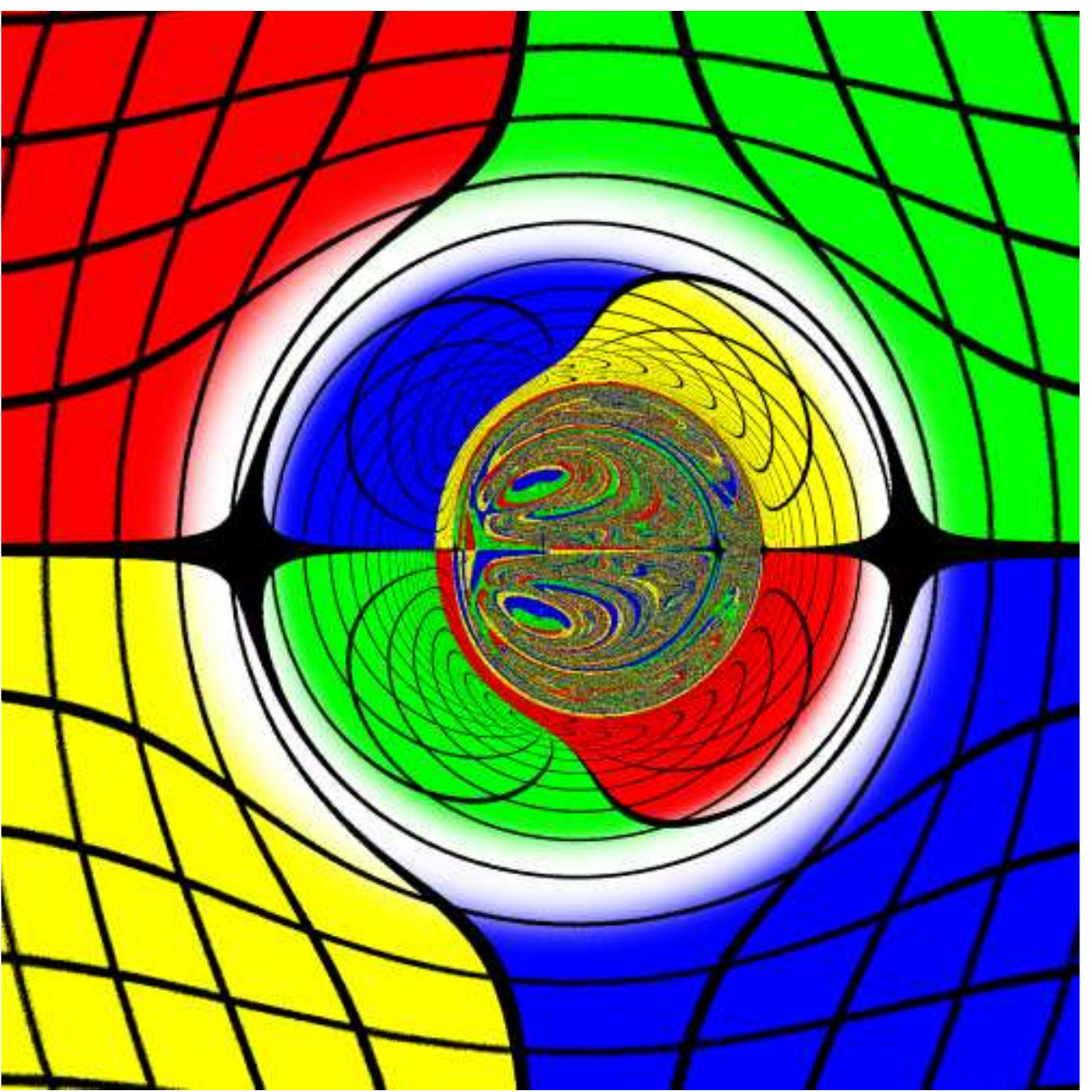}
\caption{Lensing by rotating BSs with $m=1$. From left to right: (top) $w^{(b1)}_{6,7,8}=0.95; 0.85; 0.8$; (middle) $w^{(b1)}_{9,10}=0.75; 0.7$; (bottom): $w^{(b1)}_{11}=0.65$; $w^{(b2)}_{12}=0.7$.}
\label{RBS}
\end{center}
\end{figure}

\noindent{\bf {\em Lensing by rotating BSs.}}
We now turn to rotating BSs (spin axis pointing up). Starting from vacuum, we again find a region of \textit{non-compact} BSs, $i.e$ without multiple images. Two differences, however, with respect to the top left panel of Fig.~\ref{SBS}  are: an asymmetric lensing, with an amplification of the side rotating away from $\mathcal{O}$ and the slight shift of point $F$ to the left, due to frame dragging; these are illustrated in Fig.~\ref{RBS} for a BS with $w^{(b1)}_{6}=0.95$. At $w^{(b1)}_{ER1}\simeq 0.92$ an Einstein ring appears, starting the set of \textit{compact} rotating BSs. This is the case, in Fig.~\ref{RBS}, for the BSs with $w^{(b1)}_{7,8}=0.85; 0.8$. The ring encloses again two inverted copies of part of the celestial sphere, but it is now elliptic and the duplicated image of the side rotating towards the observer is suppressed. The inversion shifts point $F$ to the right.

Further following the spiral, new Einstein `rings' appear, just as in the case of spherical BSs, except that instead of an `O-shape', they have a squashed `D-shape'. An example, for a BS with $w^{(b1)}_{9}=0.75$ is shown in Fig.~\ref{RBS}. Then, a light ring on the equatorial plane appears at $w^{(b1)}_{LR}\simeq 0.747$. Beyond this point multiple (presumably infinitely many) images of the celestial sphere arise, with, we conjecture, a fractal structure. This is illustrated by the BSs with  $w^{(b1)}_{10,11}=0.7;0.65 $ and $w^{(b2)}_{12}=0.7$ in Fig.~\ref{RBS}.

\begin{figure}[h!]
\begin{center}
\includegraphics[width=0.155\textwidth]{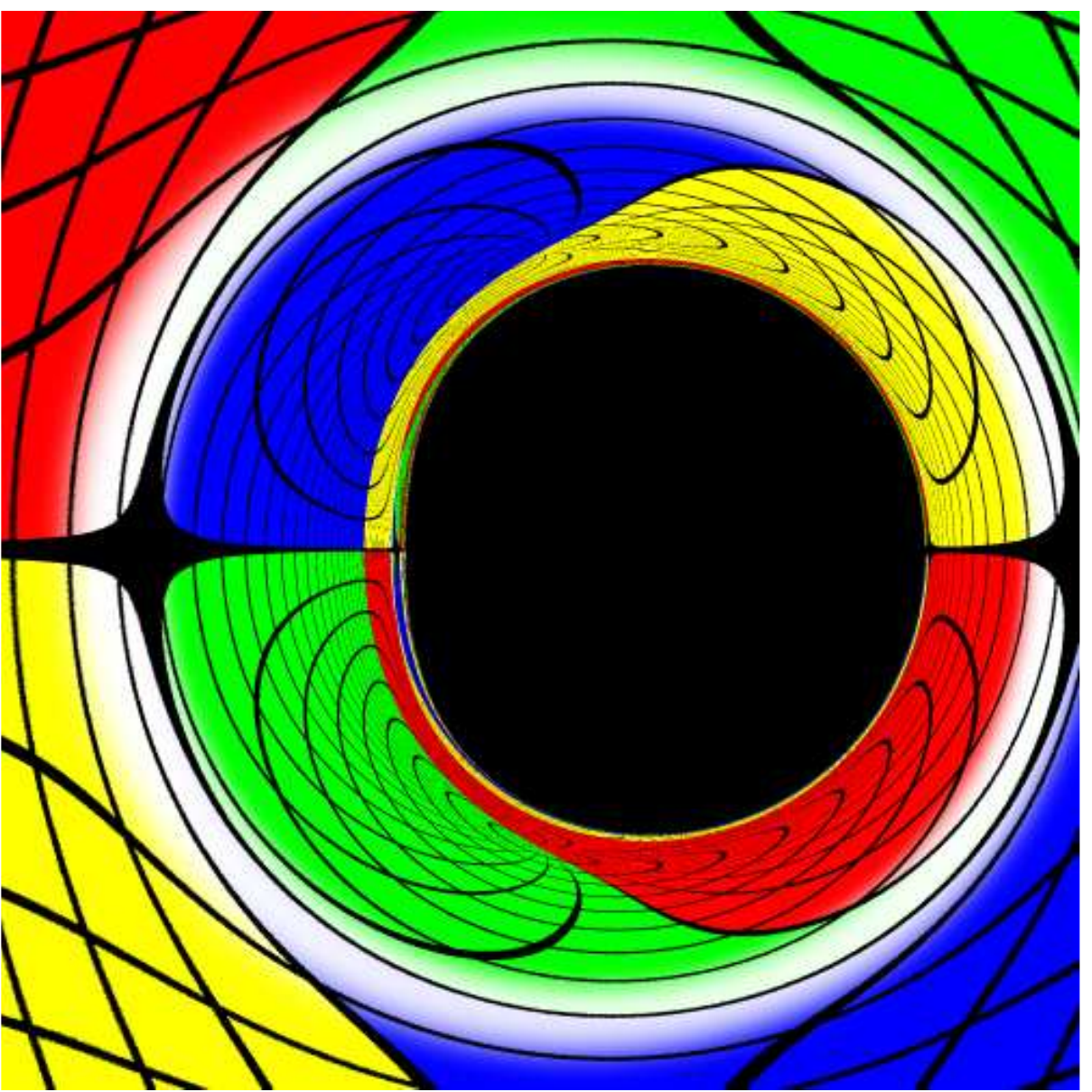}
\includegraphics[width=0.155\textwidth]{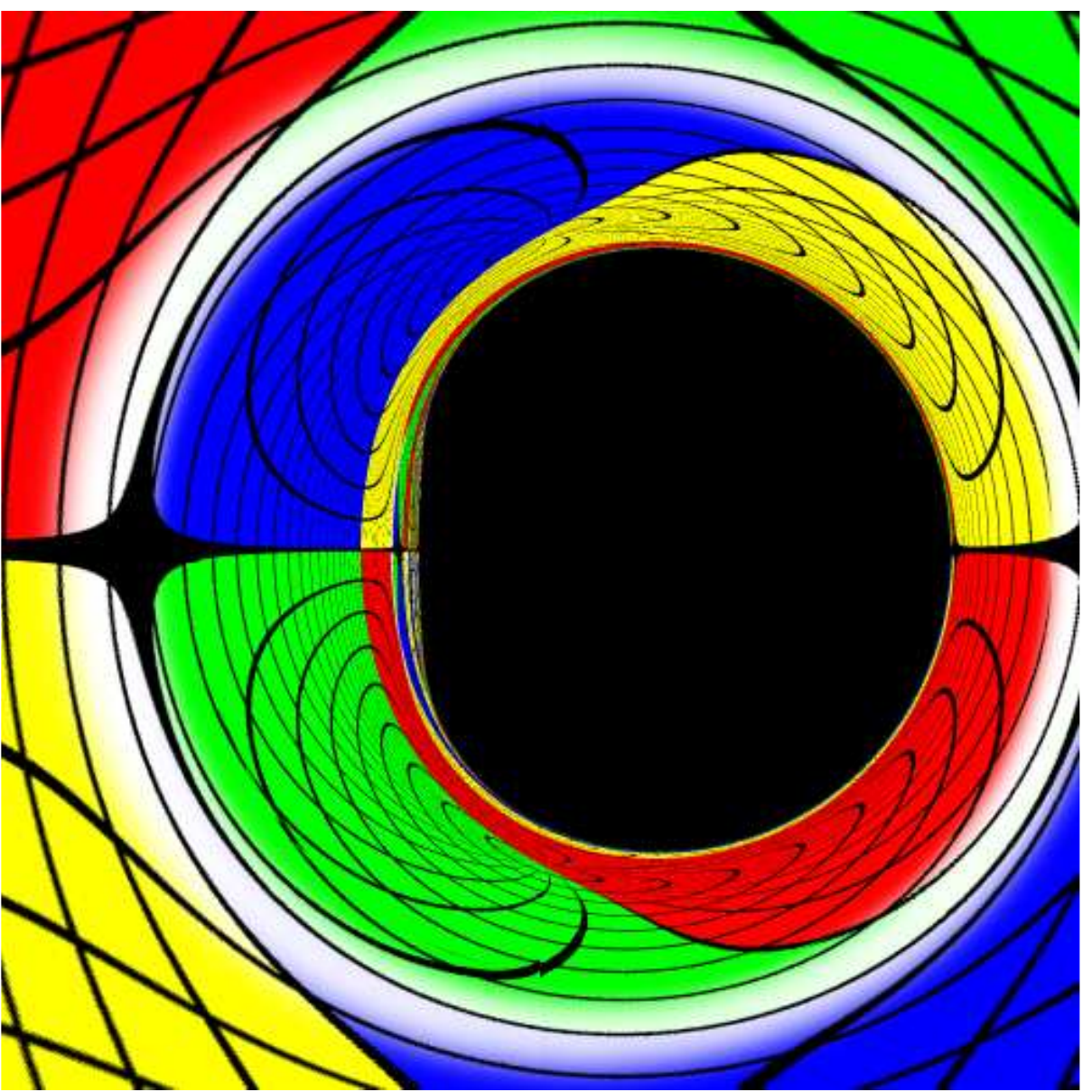}
\includegraphics[width=0.155\textwidth]{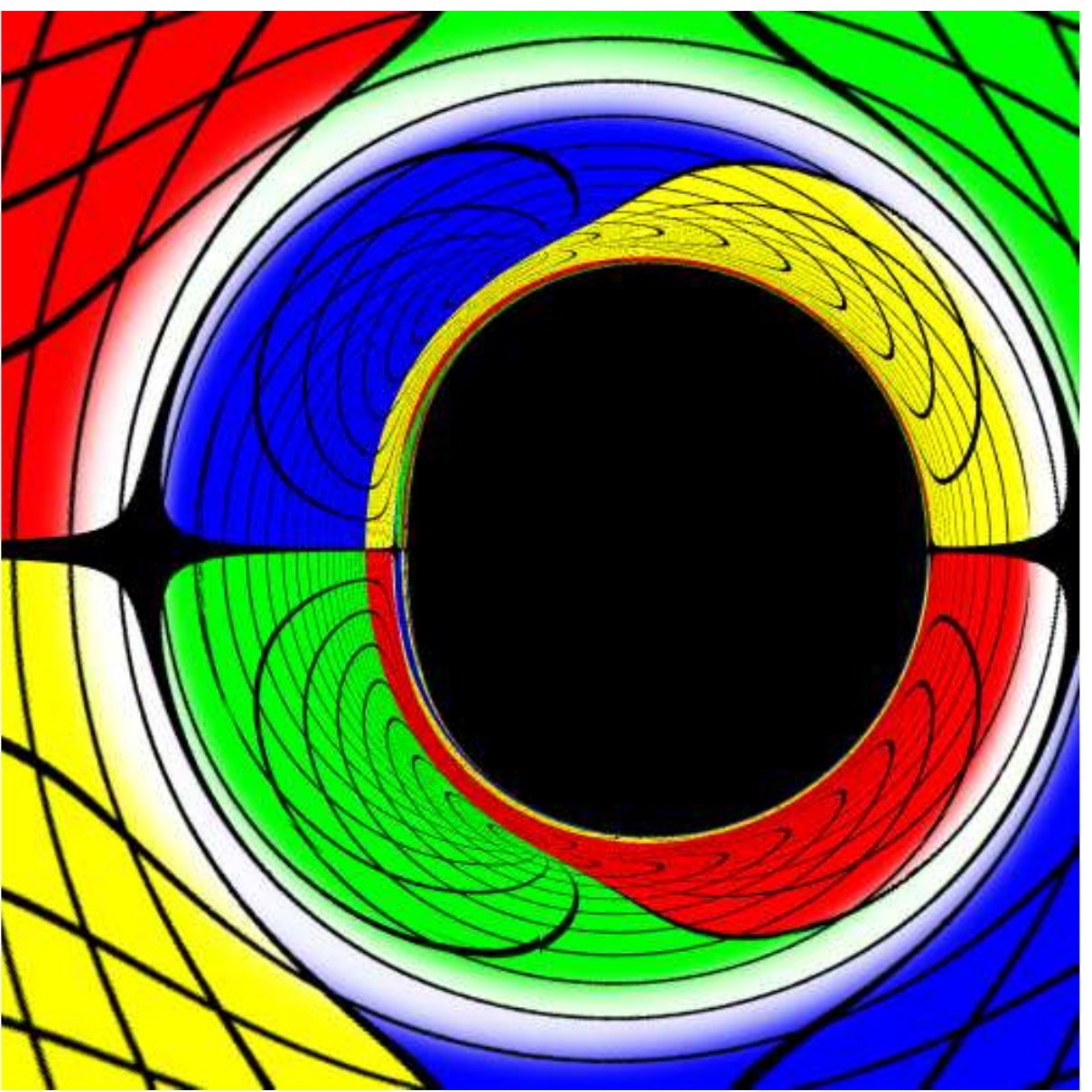}
\includegraphics[width=0.236\textwidth]{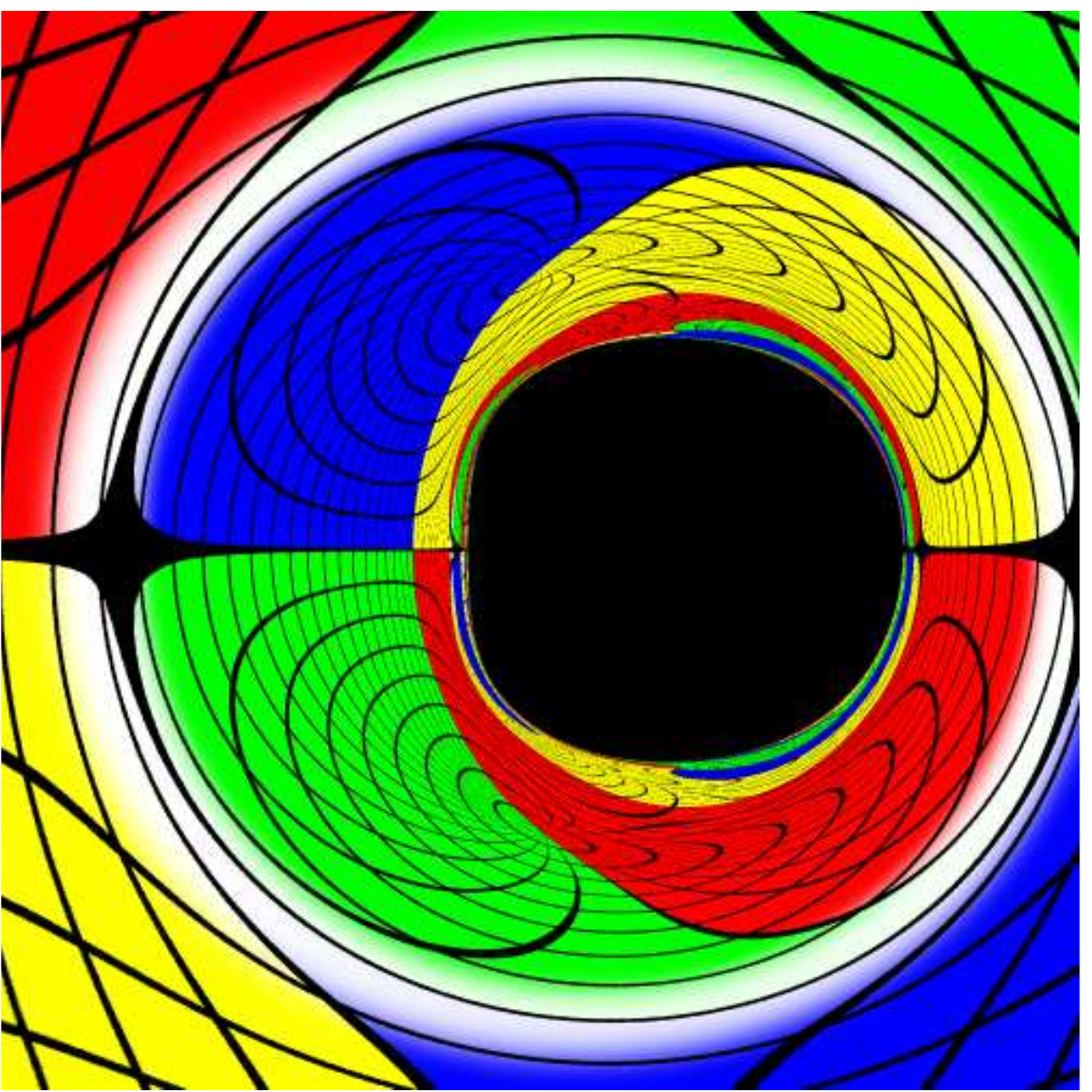}
\includegraphics[width=0.236\textwidth]{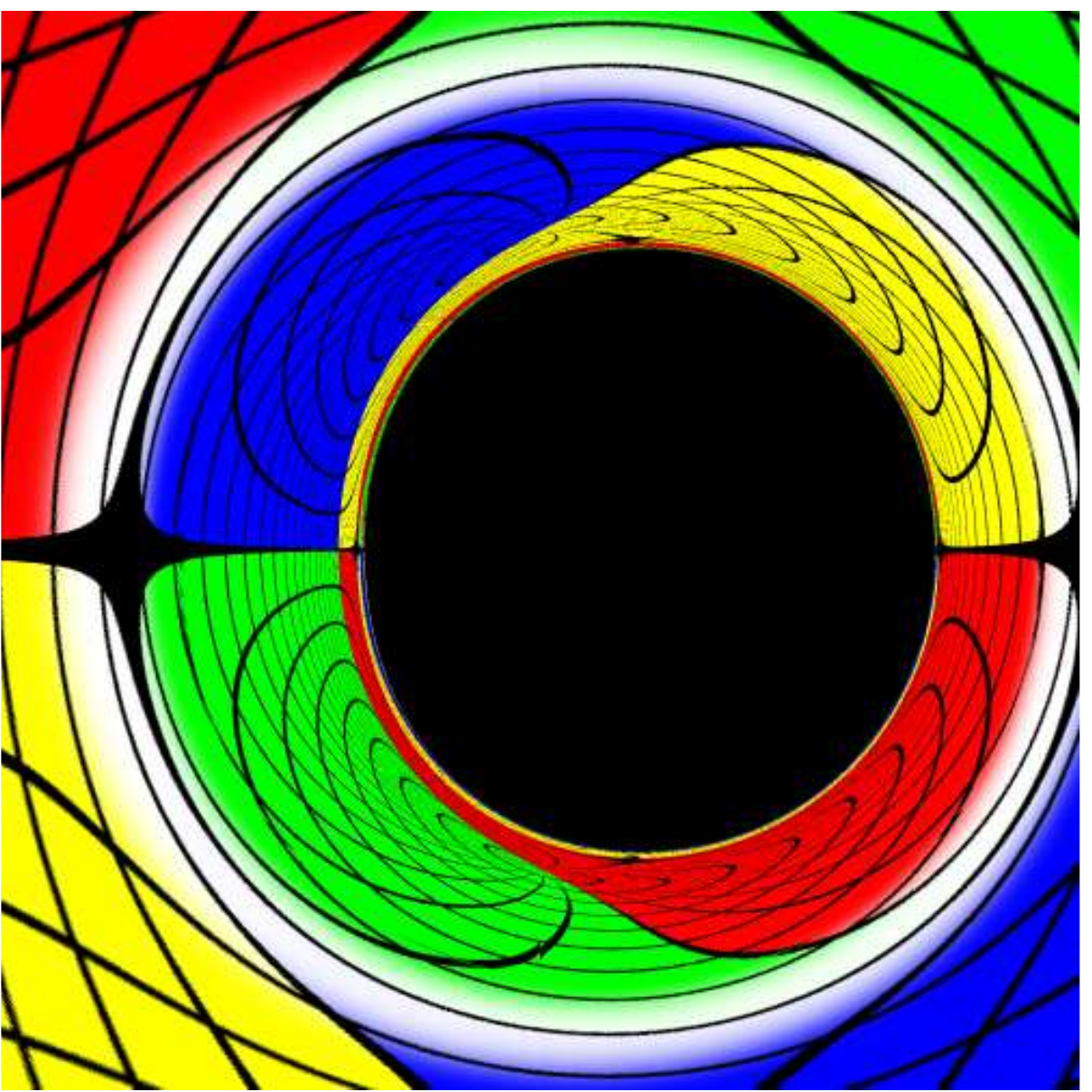}
\includegraphics[width=0.115\textwidth]{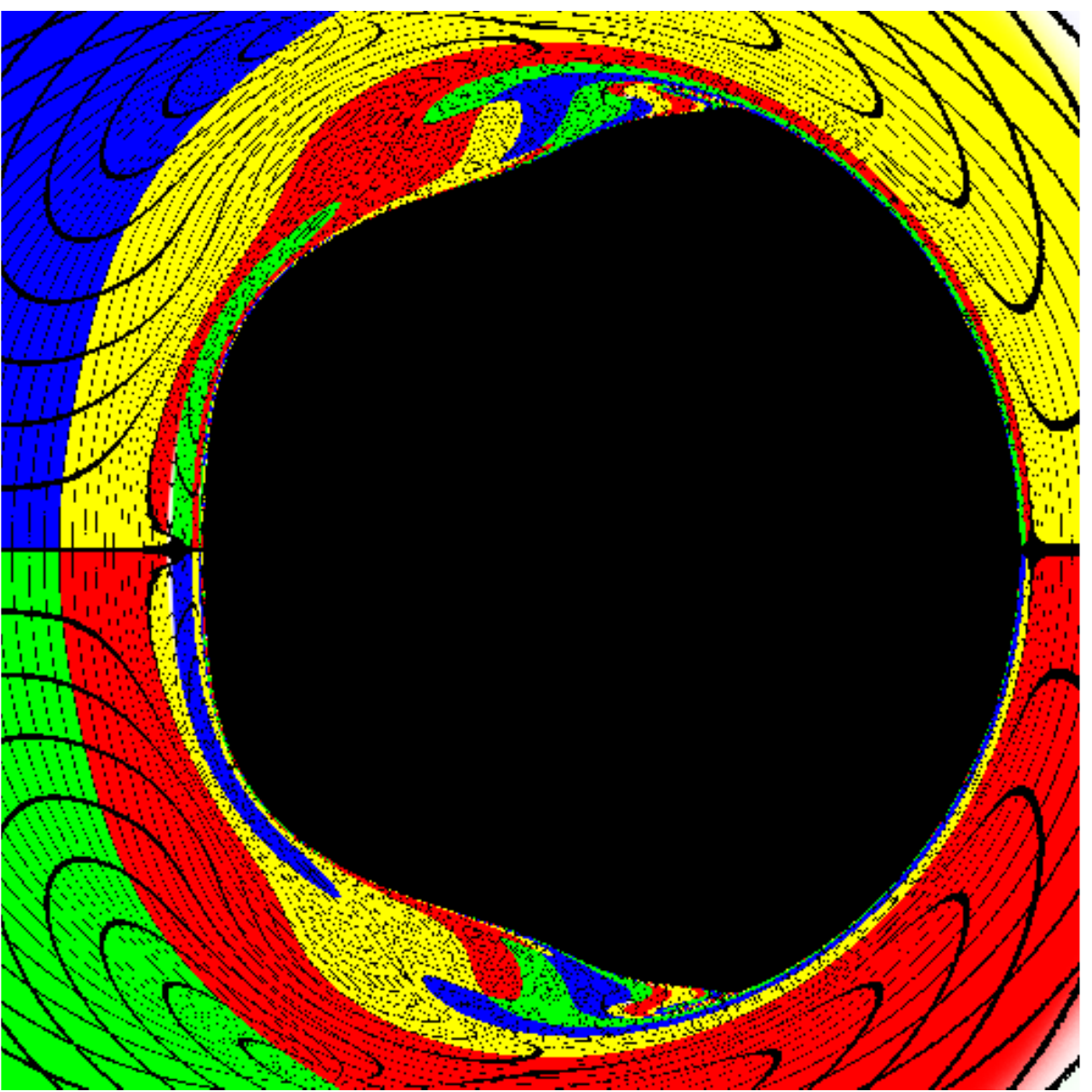}
\includegraphics[width=0.115\textwidth]{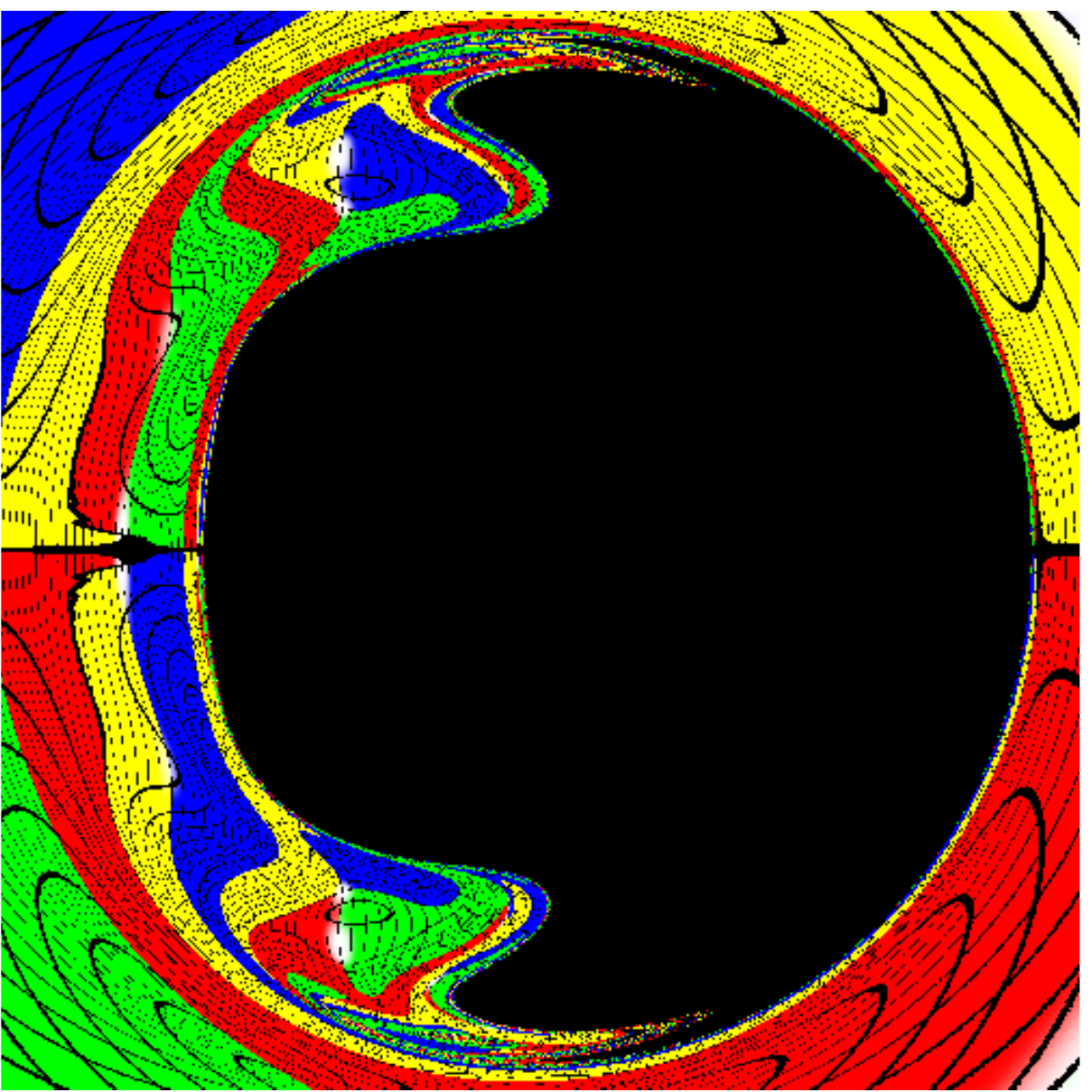}
\includegraphics[width=0.115\textwidth]{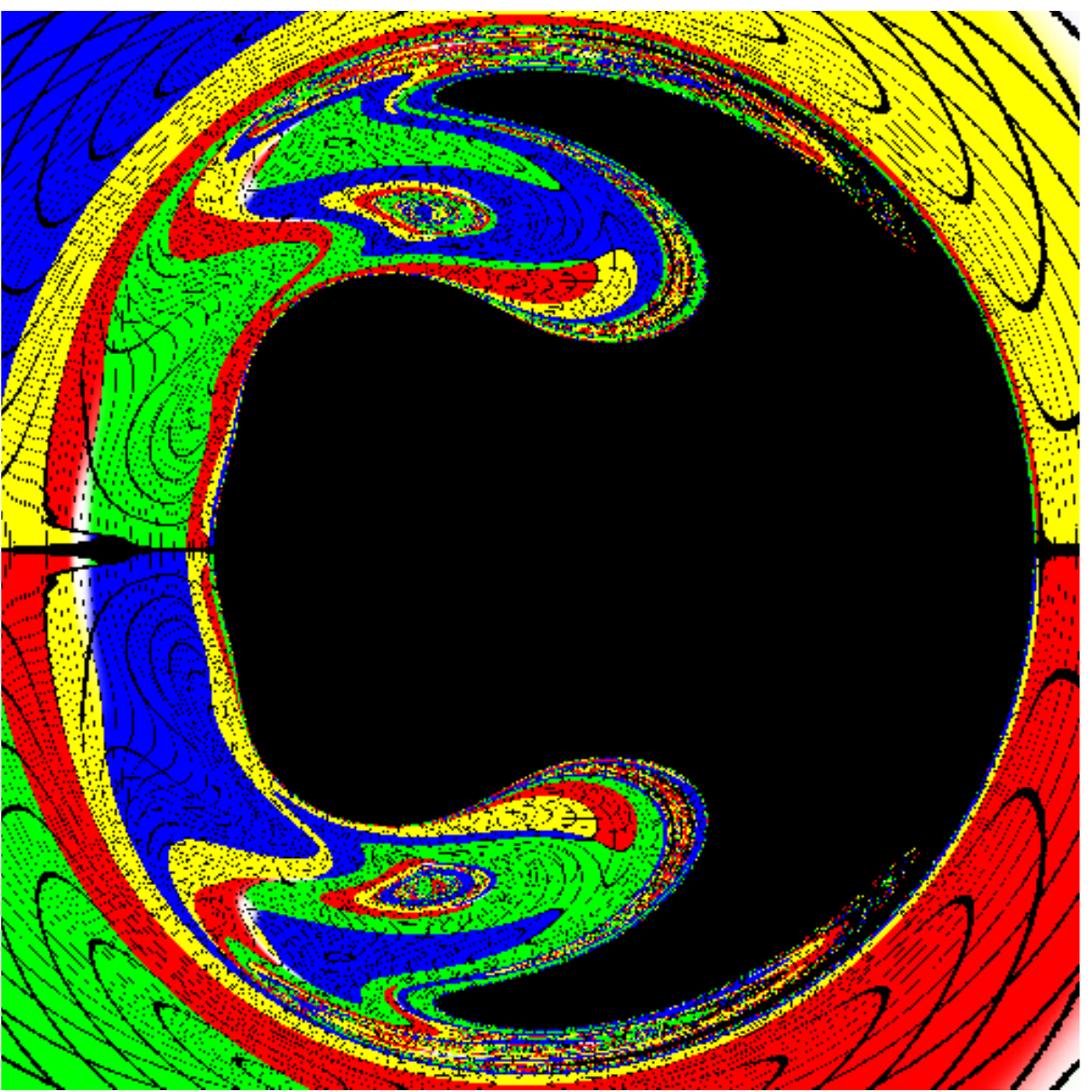}
\includegraphics[width=0.115\textwidth]{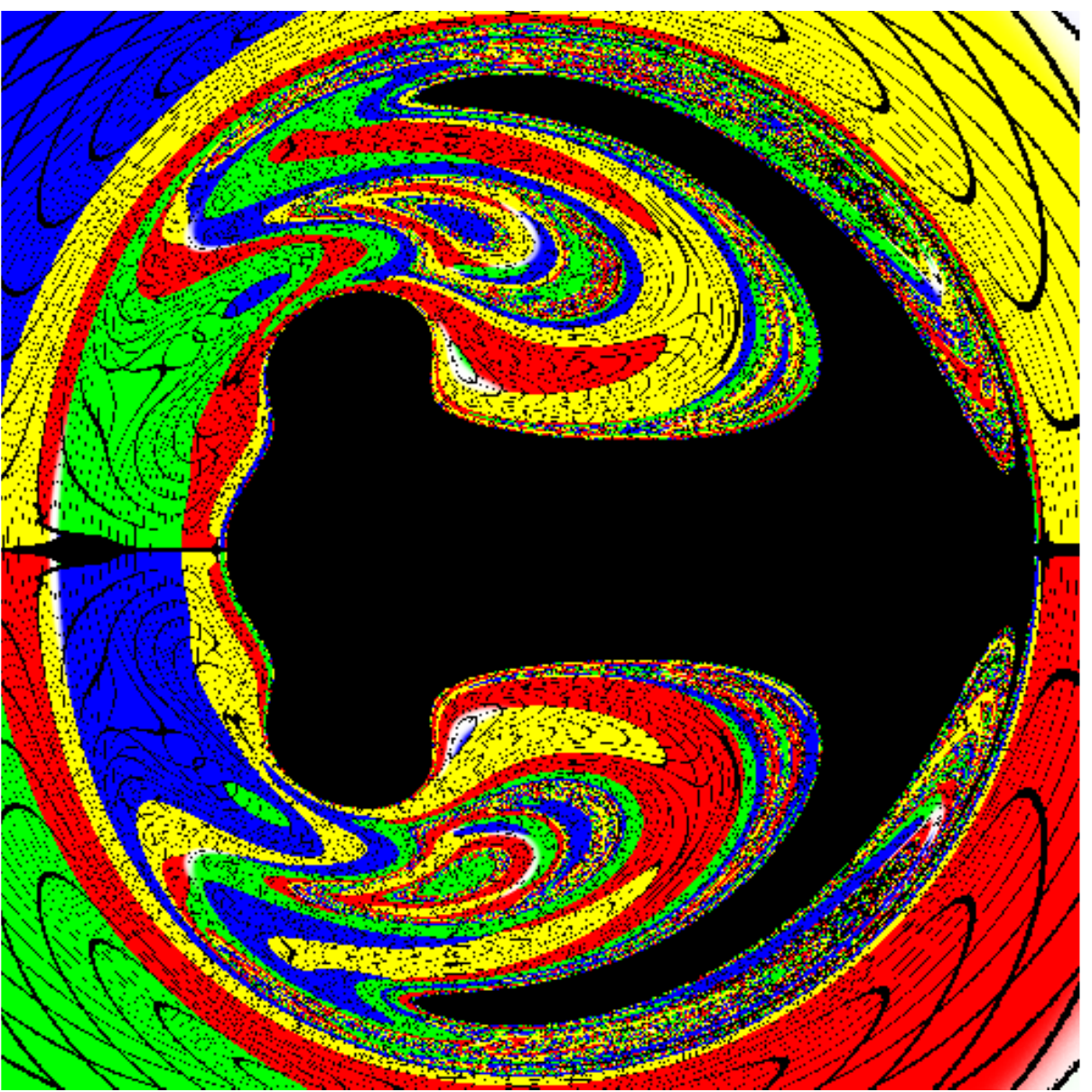}
\includegraphics[width=0.236\textwidth]{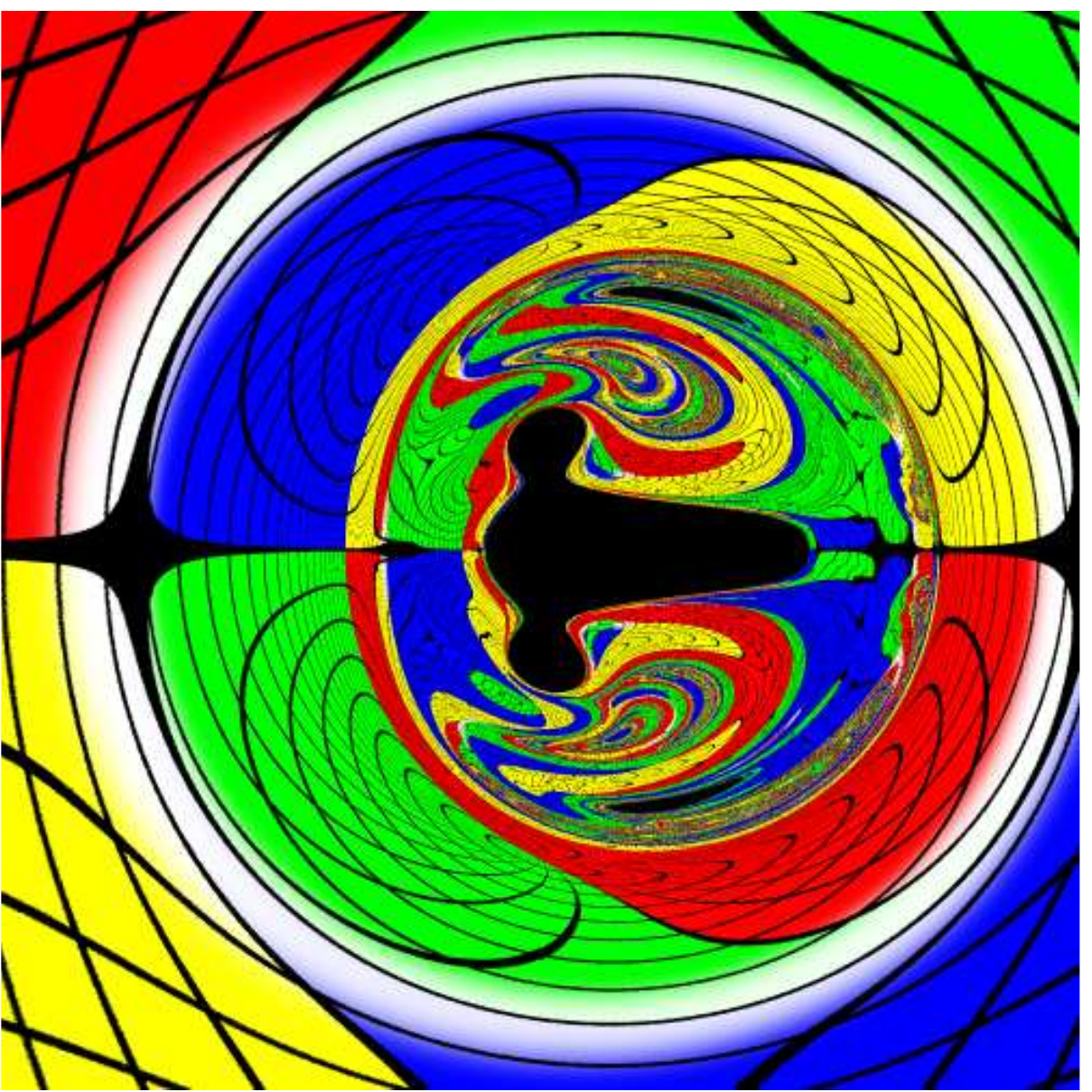}
\includegraphics[width=0.236\textwidth]{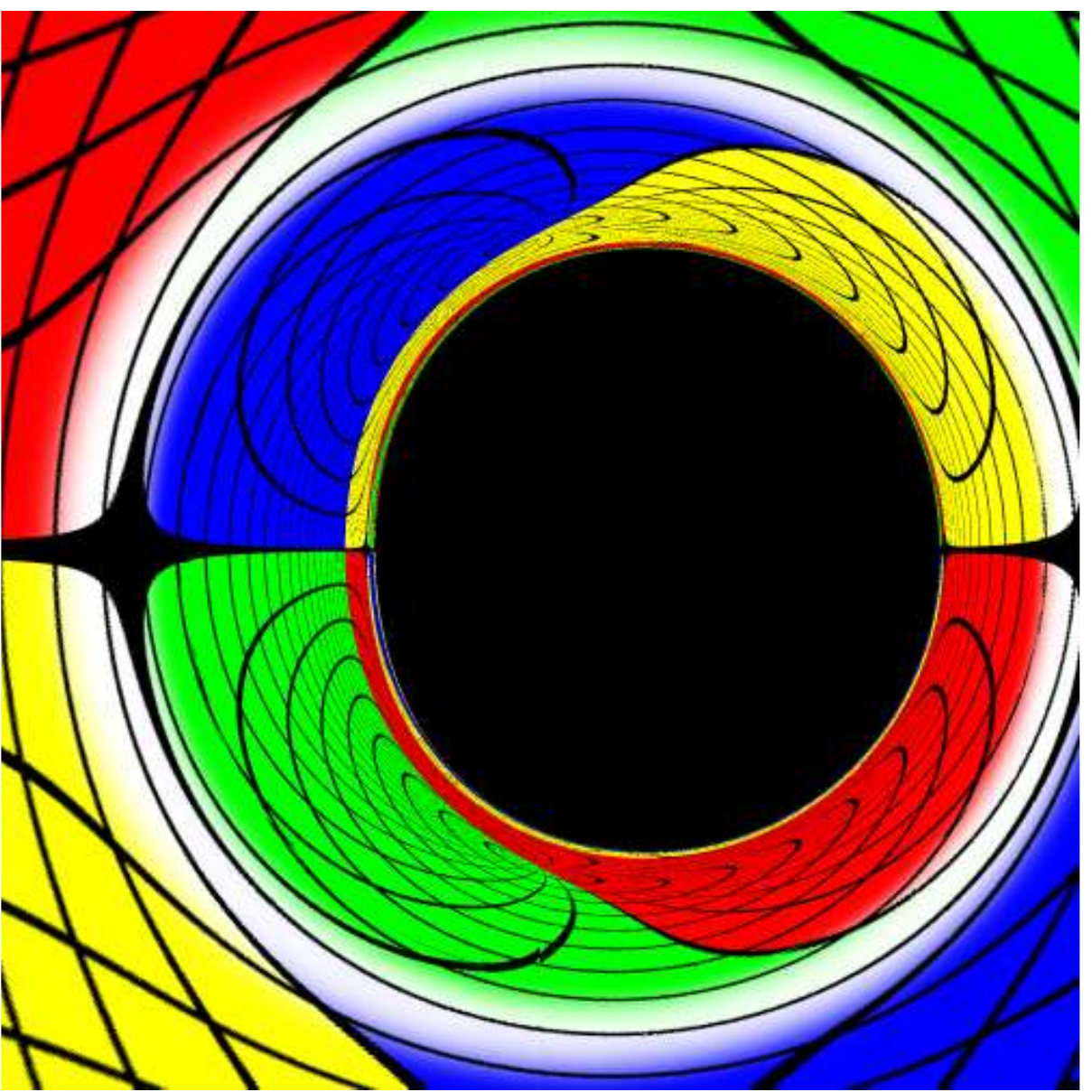}
\caption{From left to right: shadows of (top) configuration I, Kerr$_{\rm =ADM}$ and Kerr$_{\rm =H}$;  (2$^{\rm nd}$ row) configuration II and Kerr$_{\rm =ADM}$; (3$^{\rm rd}$ row) transition configurations between II and III (detail);  (bottom) configuration III and Kerr$_{\rm =ADM}$.}
\label{shadows}
\end{center}
\end{figure}

\noindent{\bf {\em Shadows of KBHsSH.}} 
A KBHSH may be regarded as a BS around (and co-rotating with) a central horizon. The latter may be \textit{non-Kerr-like}, $i.e$ violate the Kerr bound in terms of horizon quantities~\cite{Herdeiro:2015moa} and the former may have strong lensing effects, as seen above. Consequently, it is expectable that KBHsSH with ultra-compact BS-like hair and with a non-Kerr-like central horizon will have unfamiliar shadows. This expectation is confirmed in Fig.~\ref{shadows},  where we exhibit the shadows and lensing for three examples of KBHsSH (together with four `transition' examples) and for comparable Kerr BHs. The three main examples are configuration I-III in Fig.~\ref{spaceofsolutions}, and their physical quantities, in units of $\mu$ when dimensionful, are summarized as follows:

\begin{center}
\begin{tabular}{l*{7}{c}r}
\hline
  & $M_{\rm ADM}$ & $M_{\rm H}$ & $J_{\rm ADM}$ & $J_{\rm H}$ & $ \frac{M_{\rm H}}{M_{\rm ADM}}$ & $\frac{J_{\rm H}}{J_{\rm ADM}}$&$\frac{J_{\rm ADM}}{M_{\rm ADM}^2}$ & $\frac{J_{\rm H}}{M_{\rm H}^2}$\\
\hline
 I  & 0.415 & 0.393 & 0.172 & 0.150 & 95\% & 87\% & 0.999 & 0.971\\
 \hline
 II  & 0.933 & 0.234 & 0.740 & 0.115 & 25\% & 15\% & 0.850 & 2.10\\
 \hline
 III  & 0.975 & 0.018 & 0.85 & 0.002 & 1.8\% & 2.4\% & 0.894 & 6.20\\
 \hline
\end{tabular}\\
\end{center}

 In Fig.~\ref{shadows}, we have scaled $\mu$ in each case so that $M_{\rm ADM}$ is the same for all KBHsSH presented, and $\mathcal{O}$ sits at $R=15M_{\rm ADM}$. Comparing the shadows in this way bears more significance for real observations, where $M_{\rm ADM}$ of the BH is fixed by the data.

The top left panel of  Fig.~\ref{shadows} shows the shadow of configuration I (shadow I for short). For this  KBHSH the `hair' contains only $5\%$ of the total mass; the `nearby' BSs in Fig.~\ref{spaceofsolutions} are non-compact; and the central horizon is Kerr-like ($J_{\rm H}/M_{\rm H}^2<1$). The shadow, albeit qualitatively familiar, is nevertheless distinguishable from that of the Kerr$_{\rm =ADM}$ BH, which is exhibited in the top middle panel. The latter is slightly larger and more $D$-like -- a characteristic of extremal Kerr BHs. Shadow I  turns out to be closer to the one of the Kerr$_{\rm =H}$ BH, exhibited in the top right panel. This observation can be quantitatively checked: $\sigma_{\rm K}=4.81\% / 0.52\%$, taking Kerr$_{\rm =ADM}$/Kerr$_{\rm =H}$ for the comparable BH, $cf.$ Table~I.

New types of BH shadows, quite distinct from those of Kerr$_{\rm =ADM}$ BHs, appear on the left 2$^{\rm nd}$ and 4$^{\rm th}$ row panels of Fig.~\ref{shadows}, corresponding to shadow II and III. In both cases, the central BH is non-Kerr-like ($J_{\rm BH}/M_{\rm BH}^2>1$). Shadow II is smaller (average radius of $\sim$75\%)  than that of the Kerr$_{\rm =ADM}$ BH. It is also more `square', with a larger normalized deviation from sphericity. Shadow III is remarkably distinct. Its central BH has $J_{\rm H}/M_{\rm H}^2\sim 6$, allowed by the `heavy' hair that it is dragging ($cf.$ the discussion in~\cite{Herdeiro:2015moa}). The lensing of this hair resembles closely that of the ultra-compact BS on the bottom left panel of Fig.~\ref{RBS}. Interestingly, \textit{multiple} disconnected shadows of the (single) BH occur: the largest ones (besides the main `hammer-like' shadow) are two \textit{eyebrows}~\cite{Yumoto:2012kz,Abdolrahimi:2015rua}, at symmetric positions above and below the main shadow; but we have detected many other smaller shadows, hinting again at a self-similar structure. On the  3$^{\rm rd}$ row of Fig.~\ref{shadows}, the shadows of four solutions in between configurations II and III illustrate the transition between them.  Finally, we remark that the shadows of KBHsSH can have arbitrarily small sizes by considering solutions arbitrarily close to the BS curve in Fig.~\ref{spaceofsolutions}.

\begin{table}
\begin{center}
\begin{tabular}{l*{6}{c}r}
\hline
   & $D_C$ & $D_{x}$ & $D_{y}$ & $\bar{r}$ & $\sigma_r$ & $\frac{\sigma_r}{\bar{r}}$(\%)&$\sigma_{\textrm{K}}$(\%) \\
\hline
Shadow I  & 2.07& 8.48& 9.33& 4.48& 0.170& 3.8& 4.81/0.52\\
 Kerr$_{\rm =ADM}$  &2.30& 8.73& 9.76 & 4.67& 0.217& 4.64& 1.21\\
 Kerr$_{\rm =ADM}$ A & 2.38& 8.66& 9.86& 4.70& 0.260& 5.54& 0\\
 Kerr$_{\rm =H}$ A & 2.07& 8.48& 9.36& 4.50& 0.180& 3.99& 0\\
 \hline
Shadow II  & 2.39& 7.14& 6.93& 3.60& 0.118& 3.29& 25.5\\
 Kerr$_{\rm =ADM}$ A & 1.79& 9.32& 9.86& 4.82& 0.103& 2.15& 0\\
 \hline
Shadow III  & 1.79& 5.30& 4.67& 1.63& 0.838& 51.3& 68.1\\
 Kerr$_{\rm =ADM}$ A & 1.92& 9.22& 9.86& 4.80& 0.125& 2.60& 0\\
 \hline
\end{tabular}\\
\bigskip
 \caption{Parameters for Kerr shadows with `$A$' are computed from the analytic solution~\cite{1973blho.conf..215B}; $\sigma_{\textrm{K}}$ is always computed with respect to such solution. The second line in the table, computed for a Kerr BH generated numerically and using the same ray tracing code as for KBHsSH, estimates the numerical error $(\sim 1\%)$ of the KBHsSH shadows.}
\end{center}
\end{table}

\noindent{\bf {\em Remarks.}}
KBHsSH can lead to qualitatively novel types of shadows in GR, as shown by shadows II and III. Even for KBHsSH close to Kerr, their shadows are distinguishable from the latter, with the same asymptotic quantities, as illustrated by shadow I. Regardless of the astrophysical relevance of these solutions -- which is unclear -- they can yield new templates with small or large deviations from the Kerr shadows, hopefully of use for VLBI observations.  An exhaustive analysis of KBHsSH shadows spanning the space of solutions in~Fig.~\ref{spaceofsolutions}, and at different observation angles, will be presented elsewhere, for producing such templates~\cite{CHRR}. But the examples herein already raise a challenge to the parameterizations of deviations from Kerr suggested in the literature~\cite{Johannsen:2011dh,Johannsen:2015pca,Johannsen:2013rqa,Cardoso:2014rha,Rezzolla:2014mua}: can they describe shadows with such large deviations?

Besides the peculiar shape of some of the shadows exhibited, this model has one general prediction: \textit{smaller observed shadows} than those expected for Kerr BHs with the same asymptotic charges. Indeed, a `smaller' central BH seems a natural consequence of the existence of hair, carrying part of the total energy. 

Finally, for the setup herein, the redshift, which depends only on the source's and ${\mathcal{O}}$'s positions, is constant throughout the image and has been neglected. 

\newpage

%%%%%%%%%%%%%%%%%%%%%%%%%%%%%%%%%%%%%%%%%%%%%%%%%%%%%%%%%%%%%%%%%%%%%
\noindent{\bf {\em Acknowledgements.}}
%\noindent{\bf{\em Acknowledgements.}}
%%%%%%%%%%%%%%%%%%%%%%%%%%%%%%%%%%%%%%%%%%%%%%%%%%%%%%%%%%%%%%%%%%%%%
We would like to thank T. Johannsen and V. Cardoso for discussions and correspondence,  and M. M. Soares for computational assistance. C. H. and E. R. acknowledge funding from the FCT-IF programme. H.R. is supported by the grant PD/BD/109532/2015 under the MAP-Fis Ph.D. programme. This work was partially supported by the NRHEPÐ295189
FP7-PEOPLE-2011-IRSES Grant, by FCT via project No.
PTDC/FIS/116625/2010 and by the CIDMA strategic project UID/MAT/04106/2013. Computations were performed at the Blafis cluster, in Aveiro University, and at the Laboratory for Advanced Computing, University of Coimbra.

%%%%%%%%%%%%%%%%%%%%%%
%%%   REFERENCES   %%%
%%%%%%%%%%%%%%%%%%%%%%

\newpage

\bibliography{letter_shadows}

%%%%%%%%%%%%%%%
%%%   END   %%%
%%%%%%%%%%%%%%%
 
\end{document}